%% file: main.tex
\documentclass[10pt,twocolumn,letterpaper]{article}
\usepackage{iccv}          

\usepackage{graphicx}
\usepackage{adjustbox}
\usepackage{multirow, booktabs}
\usepackage{siunitx}
\usepackage{makecell}
\usepackage{amsmath}
\usepackage{amssymb}
\usepackage{booktabs}
\usepackage{graphics}
\usepackage{graphicx}
\usepackage{float}
\usepackage{calc}
\usepackage{amsfonts,bm,url,color,theorem}
\usepackage{psfrag}
\usepackage{multirow,epsfig}
\usepackage{diagbox} 
\usepackage{hhline}
\usepackage{caption}

\usepackage{pifont}

\usepackage{mathtools}

\usepackage{indentfirst}
\usepackage{colortbl}

\definecolor{bestColor}{RGB}{255, 0, 0}    
\definecolor{secondBestColor}{RGB}{0, 0, 255} 
\definecolor{thirdBestColor}{RGB}{0,165, 0}  
\definecolor{qBestColor}{RGB}{0,215, 0}  
\definecolor{input_color}{RGB}{255,240,240}      

\newcommand{\best}[1]{\textcolor{bestColor}{\textbf{#1}}}      
\newcommand{\secondBest}[1]{\textcolor{secondBestColor}{\textbf{#1}}} 
\newcommand{\thirdBest}[1]{\textcolor{thirdBestColor}{\textbf{#1}}}  

\definecolor{background_color}{RGB}{255,255,255}      
\definecolor{cnn_color}{RGB}{240,248,255}  
\definecolor{transformer_color}{RGB}{255,250,240}  
\definecolor{ours_color}{RGB}{245,255,250}  
\usepackage{colortbl}  

\definecolor{cvprblue}{rgb}{0.21,0.49,0.74}
\usepackage{hyperref}
\hypersetup{
    breaklinks,
    colorlinks,
    allcolors=cvprblue
}

\usepackage[capitalize]{cleveref}
\crefname{section}{Sec.}{Secs.}
\Crefname{section}{Section}{Sections}
\Crefname{table}{Table}{Tables}
\crefname{table}{Tab.}{Tabs.}

\begin{document}

\title{PromptHSI: Universal Hyperspectral Image Restoration with\\ Vision-Language Modulated Frequency Adaptation}

\author{Chia-Ming Lee$\textsuperscript{1}\thanks{These authors contributed equally.}$, Ching-Heng Cheng$\textsuperscript{1}\footnotemark[1]$, Yu-Fan Lin$\textsuperscript{1}$, Yi-Ching Cheng$\textsuperscript{1}$, Wo-Ting Liao$\textsuperscript{1}$, \\Chih-Chung Hsu$^\dagger$$\textsuperscript{1}$, Fu-En Yang$\textsuperscript{2,3}$, Yu-Chiang Frank Wang$\textsuperscript{2,3}$\\
Institute of Data Science, National Cheng Kung University$\textsuperscript{1}$\\
Graduate Institute of Communication Engineering, National Taiwan University$\textsuperscript{2}$ NVIDIA$\textsuperscript{3}$}

\maketitle

\begin{abstract}

Recent advances in All-in-One (AiO) RGB image restoration have demonstrated the effectiveness of prompt learning in handling multiple degradations within a single model. However, extending these approaches to hyperspectral image (HSI) restoration is challenging due to the domain gap between RGB and HSI features, information loss in visual prompts under severe composite degradations, and difficulties in capturing HSI-specific degradation patterns via text prompts. In this paper, we propose \textbf{PromptHSI}, the first universal AiO HSI restoration framework that addresses these challenges. By incorporating frequency-aware feature modulation—which utilizes frequency analysis to narrow down the restoration search space—and employing vision-language model (VLM)-guided prompt learning, our approach decomposes text prompts into intensity and bias controllers that effectively guide the restoration process while mitigating domain discrepancies. Extensive experiments demonstrate that our unified architecture excels at both fine-grained recovery and global information restoration across diverse degradation scenarios, highlighting its significant potential for practical remote sensing applications. The source code is available at \hyperlink{https://github.com/chingheng0808/PromptHSI}{https://github.com/chingheng0808/PromptHSI}.

\end{abstract}

\footnote{$^\dagger$Corresponding author, email: \tt\small{cchsu@gs.ncku.edu.tw}}
\footnote{Project page: \hyperlink{https://chingheng0808.github.io/prompthsiP/static.html}{https://chingheng0808.github.io/prompthsiP/static.html}}

\input{sections/introduction}
\input{sections/relatedworks}

\input{sections/methods}
\input{sections/experiments}
\input{sections/conclusion}



\clearpage  
\twocolumn[{%
   \renewcommand\twocolumn[1][]{#1}%
   \maketitle
   \vspace{-6mm} 
   \begin{center}
       {\LARGE \textbf{Supplementary Materials}}
       \vspace{16mm} 
       
       \centering
       \includegraphics[width=\linewidth]{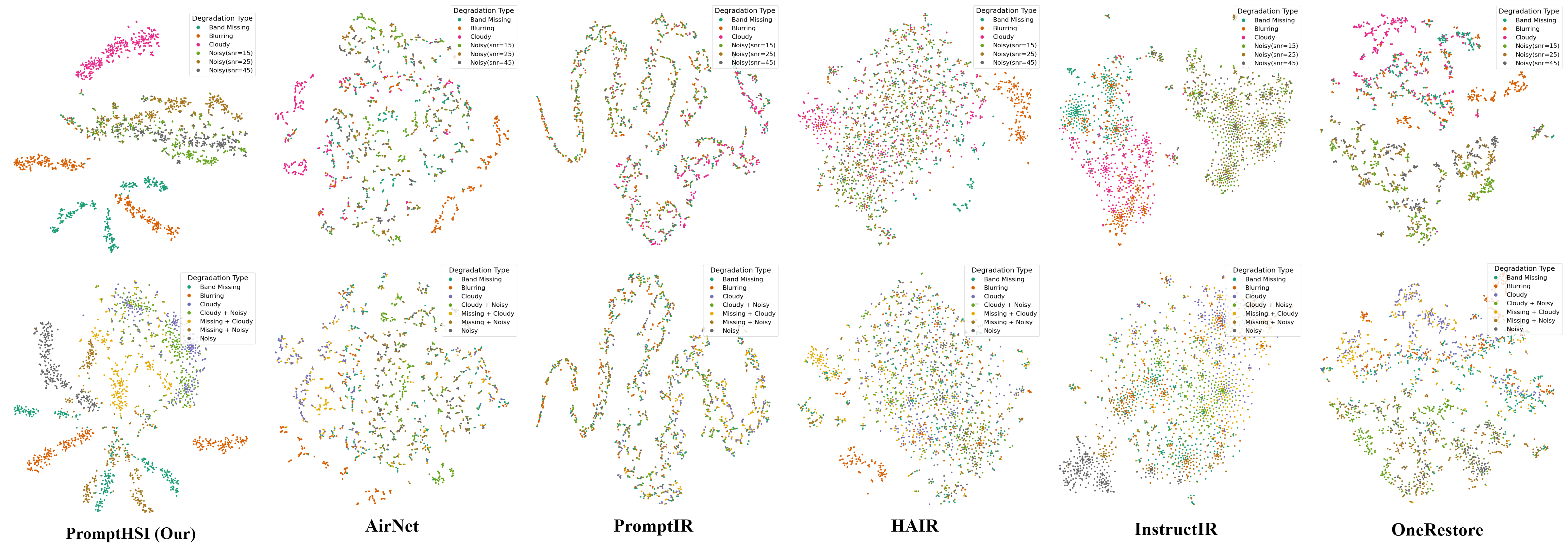}
       
       \vspace{-2mm} 
       \captionof{figure}{\small \underline{\textbf{The t-SNE map visualization for the proposed PromptHSI and other All-in-One RGR image restoration methods.}} Note that for training PromptHSI, the generated dataset includes various degradations, with each type presented separately instead of concurrently. Subsequently, we visualize the t-SNE map with \ding{182} \textbf{\emph{Top:}} multiple degradation; \ding{183} \textbf{\emph{Bottom:}} composite degradation during testing-time.}
       \label{fig:teaser}
       
       \vspace{6mm} 
   \end{center}%
}]

\appendix



\input{sections/overview_sup}

\input{sections/dataset_sup}

\input{sections/architecture_sup}

\input{sections/experiments_sup}


\clearpage
\clearpage
{
    \small
    \bibliographystyle{ieeenat_fullname_iccv}
    \bibliography{main}
}


\end{document}

%% file: sections/introduction.tex
\section{Introduction}
\label{sec:intro}

\begin{figure}
		\centering
		\includegraphics[width=0.45\textwidth]{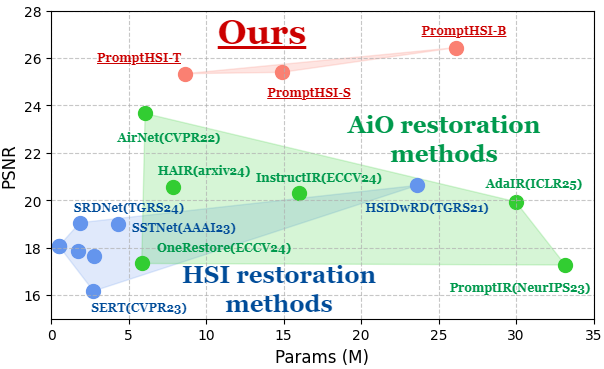}
        \vspace{-2mm}
		\caption{Performance comparison between All-in-One and HSI restoration methods on the AVIRIS dataset. The proposed \textbf{PromptHSI} offers leading performance and controllability while retaining a compact model size.
        }
	\label{fig:first}	
        \vspace{-2mm}
\end{figure}


\indent Hyperspectral images encompass a greater number of spectral bands than RGB images, offering extensive spectral details \cite{35}. Because materials reflect different spectral bands, each substance can have a unique signature for material identification \cite{HSI, HSI2, 35, lin2024selfsupervisedfusariumheadblight}. HSI can capture this valuable information and has broad application value and potential. However, because of the limitation of the device and bandwidth, observed HSIs are usually damaged by different types of degradation from camera sensing and transmission. Thus, HSI restoration is a crucial pre-processing step to recover degraded pixels from flawed observations. 

To address these issues, low-rank tensor models \cite{LRTV,10354352} are widely used in hyperspectral image restoration, relying on the assumption that the clean image $\mathbf{X}$ exhibits low-rank properties \cite{DHP,Prior1} across spatial and spectral dimensions. The restoration process is often formulated as $\mathbf{X} \approx \mathbf{L}+\mathbf{S}+\mathbf{N}$, where $\mathbf{L}$ represents the underlying low-rank structure of the clean image, $\mathbf{S}$ denotes structured sparse noise (e.g., stripe noise or dead pixels), $\mathbf{N}$ represents random noise, such as Gaussian or Poisson noise. However, under multiple degradations, this assumption often fails. Structured noise components $\mathbf{S}$ introduce high-rank artifacts, increasing the observed rank:
$\operatorname{rank}(\mathbf{X}) \gg \operatorname{rank}(\mathbf{L})$, which makes low-rank approximations less accurate. Furthermore, mixed noise complicates the separation between the clean component $\mathbf{L}$ and noise components $\mathbf{S}$ and $\mathbf{N}$, often leading to over-smoothing:
\begin{equation}
    \min_{\mathbf{L}} \| \mathcal{P}_{\Omega}(\mathbf{X} - \mathbf{L}) \|_F^2 + \lambda \operatorname{TV}(\mathbf{L}),
\end{equation}
where $P_{\Omega}$ is a projection operator on observed pixels, and $\operatorname{TV}(\mathbf{L})$ is a total variation regularization term. While this enforces spatial smoothness, it often suppresses fine details, degrading spectral fidelity. Therefore, when multiple degradations are present, the core low-rank assumption is violated, making traditional tensor-based approaches ineffective. More adaptive methods that incorporate hybrid priors \cite{ADMMADAM,FR,Fasthyin} or deep learning-based corrections are needed to achieve robust restoration.

The emergence of AiO RGB image restoration methods \cite{PromptIR, HAIR, TextualDegRemoval, InstructIR, OneRestore, adair} has demonstrated remarkable success in handling diverse degradations within a unified network. These methods leverage prompt learning, either through visual prompts \cite{PromptIR, HAIR, adair} or text-based prompts \cite{InstructIR, OneRestore}, to adaptively restore corrupted images. However, despite their effectiveness in RGB restoration, these AiO approaches face fundamental limitations when extended to HSI restoration. The primary challenge arises from the inherent complexity and variability of degradation patterns in HSI data, making it difficult to generalize the same prompt-based strategies.

To address these challenges, we propose \textbf{PromptHSI}, a frequency-aware, prompt-guided HSI restoration framework that overcomes the limitations of existing AiO RGB-based restoration methods. Unlike conventional approaches that struggle with the complexity of HSI degradation patterns, PromptHSI integrates frequency-domain priors with VLM-driven prompt learning, enabling robust and adaptive restoration across diverse degradation scenarios. In summary, our key contributions are summarized as follows:

\begin{itemize}
    \item \textbf{A Frequency-Aware Prior for Efficient Restoration.}
    We introduce a structured frequency-domain analysis that models the impact of different degradations on specific frequency components. By leveraging Fourier-based priors, PromptHSI significantly reduces the restoration search space, making it easier for deep networks to learn effective recovery mappings.

    \item \textbf{Vision-Language Modulated Prompt Learning for Controllable Restoration.}
    Unlike traditional prompt-based approaches that rely on static feature modulation, PromptHSI employs VLM-guided adaptive modulation, where textual embeddings dynamically control the restoration process. This allows for more flexible and interpretable restoration outcomes, surpassing conventional AiO RGB methods in handling complex degradations.

    \item \textbf{The First Composite Degradation Dataset for Benchmarking HSI Restoration.}
    To facilitate research in HSI restoration, we construct the first standardized composite degradation dataset, incorporating cloud occlusion, blurring, noise, and spectral band loss. We believe this dataset will be publicly released to support future developments in hyperspectral image restoration.
\end{itemize}
Through extensive experiments, we demonstrate that PromptHSI consistently outperforms state-of-the-art AiO RGB methods in HSI restoration tasks while offering controllable restoration via text prompts. 

\begin{figure}
		\centering
		\includegraphics[width=0.48\textwidth]{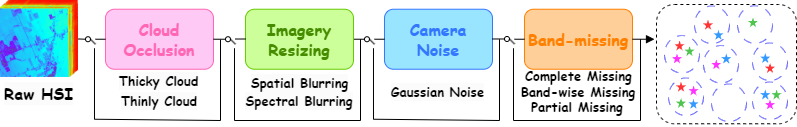}
		\caption{Illustration of the degraded HSI and description synthesis pipeline. Each degradation is activated with probability $p$ (set to 0.5 in all experiments). To ensure diversity and realism, each activated degradation randomly applies one of its sub-degradations to the input HSI. More details are provided in the \textbf{Supplementary}.
        }
		\label{fig:second}	
        \vspace{-2mm}
\end{figure}

%% file: sections/relatedworks.tex
\section{Related Works}

\subsection{Hyperspectral Image Restoration} 

Conventional approaches are based on the structural properties of HSI, such as low-rank and sparse priors \cite{DHP,Prior1}. Optimization-based methods \cite{Fasthyin,BM4D,LRTV,UBD} were initially used for HSI restoration. With the progress of deep learning, HSI can be effectively recovered by imposing appropriate constraints and suitable design for task-specific restoration. Unlike RGB image restoration, HSI restoration considers not only global information aggregation but also spatial-spectral correlation at locality.
Several works against different types of restoration model, such as denoising \cite{SST,RCTV,NGmeet} for Gaussian noise and impulse noise, inpainting \cite{UBD,Fasthyin,ADMMADAM} for stripe effect, and spatial blurring \cite{SSPSR,GDRRN,CSAKD} or limited spectral resolution \cite{SSR-dl,mst,mst_pp}. 

Despite these advancements, multiple or composite degradations simultaneously affect both structural integrity and fine-grained details in HSIs, making restoration methods highly prone to performance degradation. Moreover, erroneous compensation effects \cite{DiffBIR, AirNet} further exacerbate this issue, as models may misinterpret different degradation types, leading to incorrect adjustments. For instance, in HSIs degraded by both noise and cloud occlusion, conventional denoising methods may mistakenly treat occluded regions as random noise, resulting in spectral distortions and suboptimal restoration outcomes.




\subsection{All-in-One RGB Image Restoration} 

A recently emerging paradigm in RGB image restoration designs single models to handle multiple degradations \cite{DL,TKMANet,AirNet,PromptIR,InstructIR,HAIR,OneRestore}. These multi-task methods leverage prompting and diverse frameworks to share information between tasks and degradations. For example, AirNet \cite{AirNet} used contrastive learning to jointly learn degradation representations and restoration; PromptIR \cite{PromptIR} introduced visual prompts for degradation perception during decoding; InstructIR \cite{InstructIR} employed text prompts to guide the network; HAIR \cite{HAIR} utilized a hypernetwork to generate degradation-specific parameters; OneRestore \cite{OneRestore} extended AiO restoration to composite degradations, addressing multiple corruptions in a single image; and AdaIR \cite{adair} transforms feature maps into the frequency domain via Fourier transform and uses an adaptive mask, guided by the degraded image, to decouple low- and high-frequency features.

However, despite their effectiveness in RGB, these solutions are limited in HSIs. Due to domain gaps and composite degradation, learning a compact degradation representation and achieving effective restoration in HSIs remains challenging.

\section{Preliminary and Motivation}
\label{sec:preliminary}
\vspace*{-2mm}
\begin{figure}[h]
		\centering
		\includegraphics[width=0.48\textwidth]{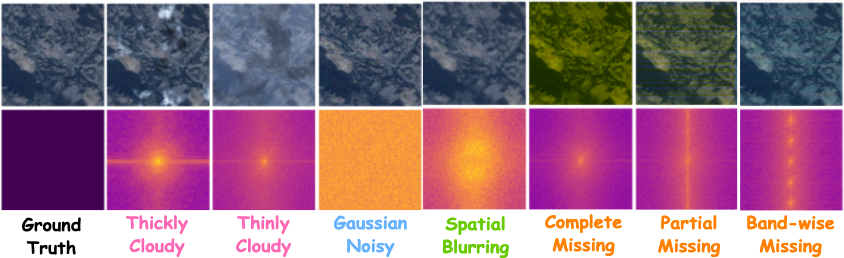}
		\caption{Visualizations of the Fourier transforms of HSIs with specific degradations, displayed from top to bottom alongside the Fourier spectra of the residual images (obtained by subtracting the degraded images from the ground truth).
        }
		\label{fig:frequency}	
        \vspace{-2mm}
\end{figure}

\subsection{Hybrid Architecture for HSI Restoration} 
Effective HSI restoration requires an architecture capable of addressing diverse degradations within a single framework \cite{Xrestormer,RUN}. Existing approaches predominantly adopt (i) U-shaped encoder-decoder networks \cite{restormer,Uformer}, which effectively capture global semantic representations for overall restoration but have difficulty with fine details, or (ii) plain residual-in-residual networks \cite{swinir,DRCT,RCAN}, which enhance spatial features and high-frequency information but lack global awareness.

AiO restoration models \cite{PromptIR,HAIR,OneRestore} and HSI-specific methods \cite{SRDNet,SNLSR,HyperRestormer} favor the former, limiting fine-detail recovery. To overcome this, PromptHSI integrates residual-in-residual blocks within a U-shaped cascade decoder, preserving global context while refining local details for greater adaptability across degradation scenarios.

\subsection{Degradations Analysis in Frequency Domain} 
\label{sec:deg}

Unlike RGB image restoration, HSI restoration must account for complex spatial-spectral correlations across hundreds of spectral bands \cite{8454887}. The degradation process can be mathematically formulated as $I_{\text{degraded}} = \mathcal{D}(I_\text{clean}) + \epsilon$, where $\mathcal{D}(\cdot)$ represents the degradation operator, encompassing effects such as blurring, stripe noise, and occlusion, while 
$\epsilon$ denotes additive noise.


\textbf{Frequency modulation.} Recovering $I_{\text{degraded}}$ from an arbitrary degradation function $\mathcal{D}^{-1}\{\cdot\}$ is inherently challenging, particularly in the presence of composite degradations, where multiple degradation factors (e.g., noise, blur, and missing bands) interact in a complex manner. Unlike isolated degradations, which can often be addressed with predefined priors, composite degradations induce coupled distortions across both spatial and spectral dimensions, making traditional inverse modeling impractical.

\begin{figure}[h]
		\centering
		\includegraphics[width=0.48\textwidth]{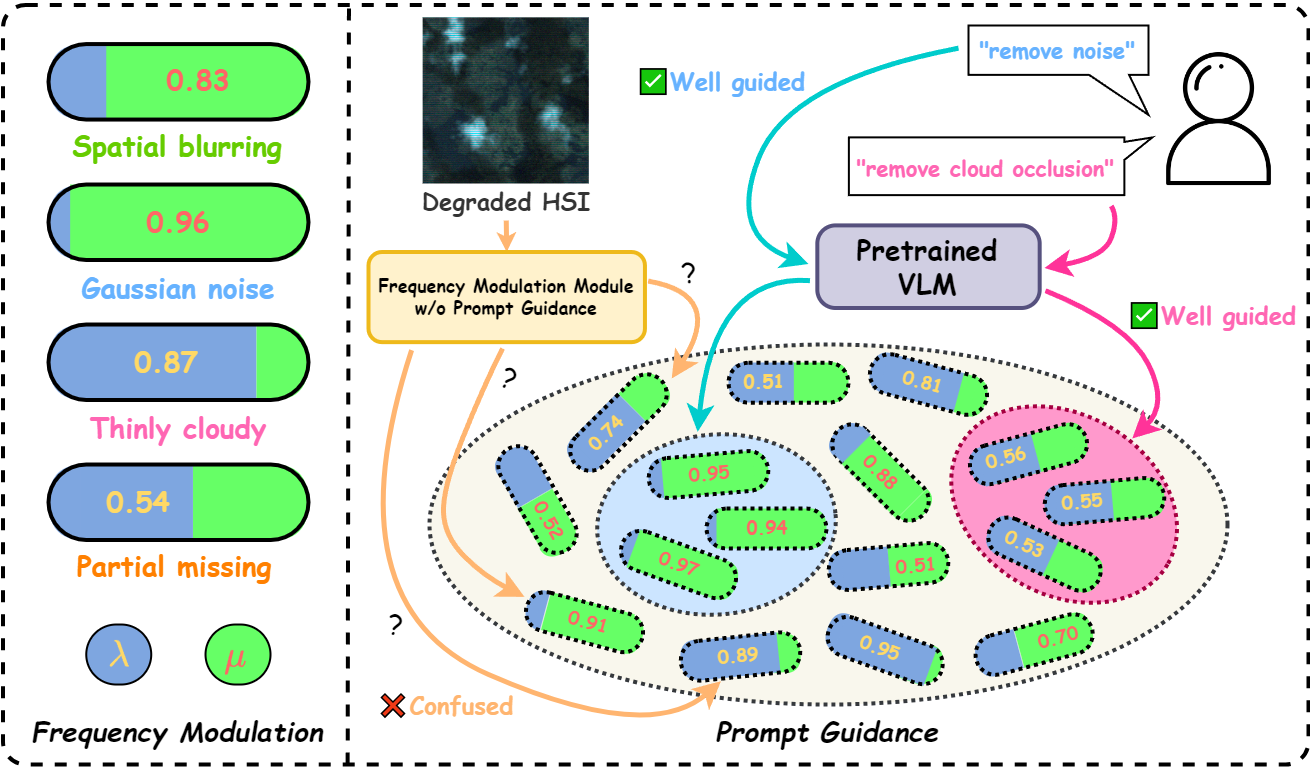}
		\caption{A diagram illustrating our proposed strategy, where integrating prompt guidance with frequency modulation facilitates the adaptive selection of $\lambda$ and $\mu$, enhancing HSI restoration under various degradation types. This mechanism encourages the model to efficiently converge toward optimal parameters.
        }
		\label{fig:freq-analysis}	
        \vspace{-2mm}
\end{figure}

Our empirical analysis (refer to Figure \ref{fig:frequency}) reveals that different types of degradation exhibit distinct but overlapping spectral characteristics. Motivated by this, we reformulate Equation \ref{eq2} by modeling the degradation process as an affine transformation in the frequency domain:
\begin{equation}
\footnotesize
F(I_{\text{degraded}}) = \hat{\lambda} \odot F(I_{\text{clean}}) + \hat{\mu} = (1+\lambda) \odot F(I_{\text{clean}}) + \mu,
\label{eq2}
\end{equation}
where $F(\cdot)$ denotes the Fourier transform, $\lambda$ modulates frequency amplitudes, $\mu$ shifts the frequency distribution, and $\odot$ represents element-wise multiplication. Under composite degradations, these frequency parameters exhibit complex dependencies:
\begin{itemize}
    \item Blurring attenuates high-frequency components  ($\lambda_{\text{blur}} \ll \mu_{\text{blur}}$), reducing fine details.
    \item Noise induces broadband perturbations, where ($\mu_{\text{noise}} \approx constant$) remains nearly uniform across frequencies.
    \item Occlusion and missing bands cause structured disruptions in both $\lambda$ and $\mu$, leading to spectral inconsistencies.
    \item Composite degradations result in frequency interactions, where overlapping effects create non-trivial distortions beyond individual degradations.
\end{itemize}
For a clearer understanding, Figure \ref{fig:freq-analysis} (left) visualizes these degradation-specific frequency characteristics, highlighting the non-additive nature of composite degradations in the Fourier domain.

\begin{figure*}
		\centering
		\includegraphics[width=1\textwidth]{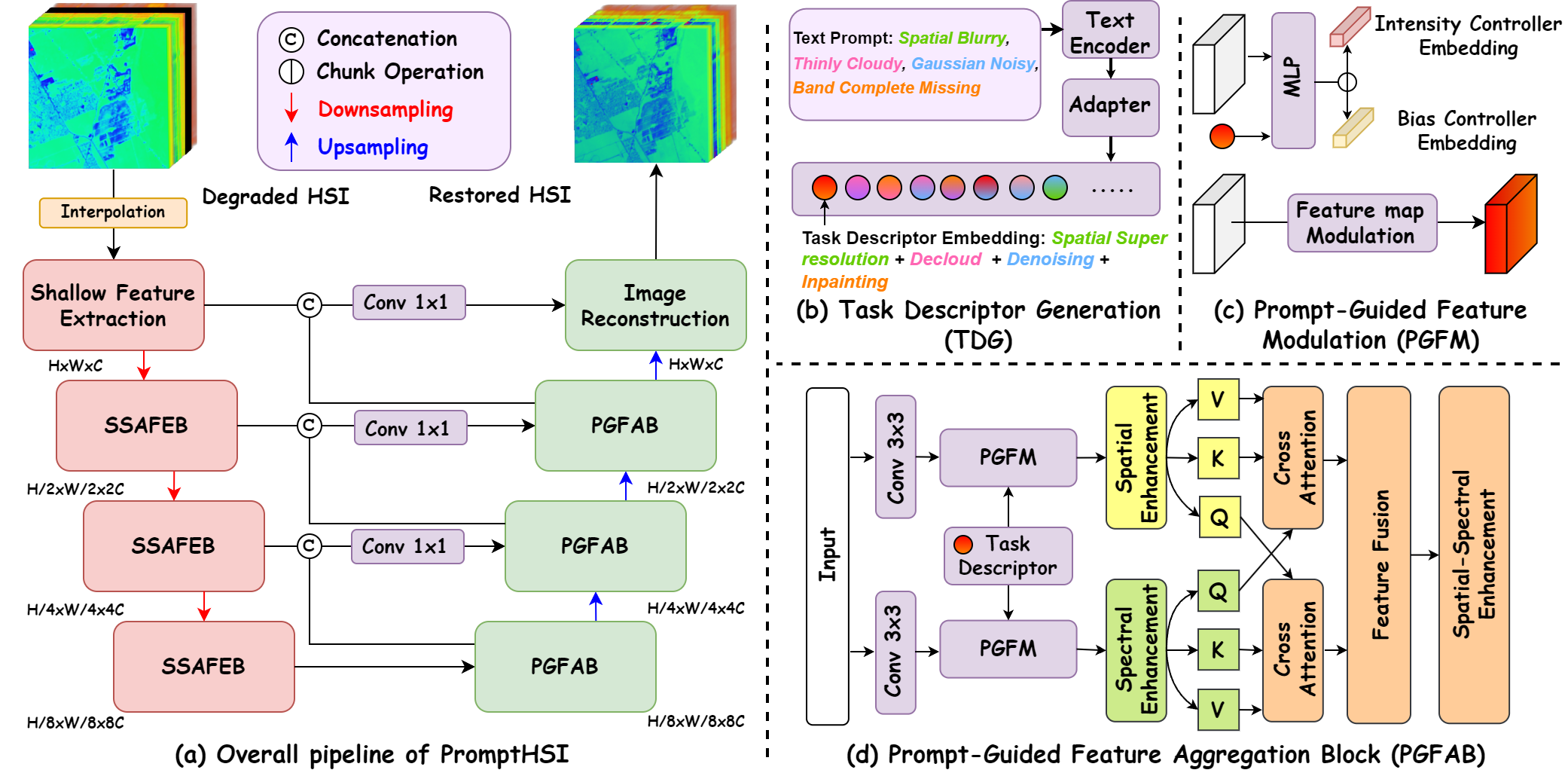}
		\caption{\small \underline{\textbf{The architecture of the proposed PromptHSI}}, which is designed to tackle the composite degradations within a single universal model via VLM-guided frequency modulation. First, we design a network based on U-net-like encoder-decoder architecture for capturing global spatial-spectral information, while combining plain-residual-in residual networks in the proposed PGFAB for spatial-spectral enhancement.
        Afterwards, text-prompt-guidance and frequency-aware feature modulation is proposed to jointly better capture the representation of composite degradation with controllability. 
  }
		\label{fig:flowchart}	
        \vspace{-2mm}
\end{figure*}

\textbf{Prompt guidance.} The diverse nature of HSI degradations presents significant challenges in learning the corresponding frequency coefficients. As shown in Figure \ref{fig:freq-analysis} (right), previous AiO restoration methods \cite{HAIR, adair} utilize adaptive modules to classify degradation types. However, lacking explicit prompt-based guidance, these methods rely solely on implicit feature learning, making parameter selection unreliable, particularly under composite degradations, where multiple degradation effects exhibit complex spectral interactions.

To address this, we introduce a VLM-driven prompt guidance mechanism, a core component of PromptHSI, which embeds frequency-aware priors into restoration. Leveraging textual cues from a given prompt $T$, our approach modulates frequency coefficients, refining the parameter space for faster convergence and enhanced frequency adaptation. This ensures precise control, adaptability, and robust performance across diverse degradations.

%% file: sections/methods.tex
\section{Proposed Method}
\label{sec:method}

\textbf{Overview.} Figure \ref{fig:flowchart} illustrates the overall architecture of PromptHSI, which consists of three key components designed to address the challenges in universal HSI restoration: \ding{182} A unified U-shape network which contains Spatial-Spectral Aware Feature Extraction Block (SSAEFB) and Propmpt-guided Feature Aggregation Block (PGFAB) that effectively captures both enhanced spatial-spectral fine-grained details and global contextual information from degraded HSI; \ding{183} A Prompt-Guided Feature Modulation (PGFM) that effectively bridging the domain gap between textual features and HSI features through frequency-domain modulation, while avoiding direct cross-domain feature alignment and information loss.

\subsection{Hyperspectral Image Restoration Pipeline}

Our restoration pipeline consists of two key components: a SSAEFB for efficient feature extraction, and a PGFAB for controllable restoration with text-prompt guidance.

The encoder exploits the low-rank and sparse prior of HSIs through depth-wise separable convolution, effectively reducing computational complexity while preserving spectral characteristics. PGFAB then utilizes dual branches for spatial and spectral enhancement, with cross-attention mechanisms facilitating comprehensive feature integration.

\textbf{Spatial-spectral-aware encoder.} 
In conventional RGB restoration and other vision tasks, standard $3\times3$ convolutions are widely considered to be sufficiently versatile and effective feature encoders. However, considering the low-rank characteristics of HSI, where substantial redundancy exists between different spectral bands, utilizing standard convolutions may reference unnecessary information from distant spectral bands, leading to suboptimal performance. Therefore, encoders based on lightweight convolutions prove sufficiently effective while further reducing parameter size to meet practical application requirements. 

Specifically, for the input HSI $I_\text{degraded}$, we first interpolate the spectrally discontinuous regions (caused by the stripe effect) to facilitate the learning process. Subsequently, we apply standard convolution layers to obtain shallow feature maps. 

Afterwards, SSAFEBs is based on a depth-wise separable convolution to extract features while inherently taking advantage of this low-rank prior, thereby reducing computational complexity without compromising performance \cite{rtcs}.


\textbf{Propmpt-guided feature aggregation block.} The features captured by SSAFEB are combined with the informative text-prompt guidance from PGFM (which will be discussed in the next subsection). Using effective guidance, the network is allowed to dynamically activate the corresponding weight for HSI restoration according to different types of degradation. To capture detailed information, PGFAB utilizes two distinct branches: one focuses on spatial dimension and the other on spectra. These branches are eventually integrated for feature aggregation. 

\textbf{Spatial and spectral feature enhancement.} The PGFM output is first refined using sequential RDG \cite{DRCT} and FRDB \cite{rtcs} modules to enhance fine details. Next, a cross-attention layer integrates spatial and spectral features from different branches. Specifically, we compute:
\begin{equation}
\footnotesize
\{Q_{\alpha}, K_{\alpha}, V_{\alpha}\} = \mathcal{F}^{\alpha}_{qkv}(F_{\alpha}),\quad \{Q_{\beta}, K_{\beta}, V_{\beta}\} = \mathcal{F}^{\beta}_{qkv}(F_{\beta}),
\end{equation}
where $F_{\alpha}$ and $F_{\beta}$ are the spatial and spectral features, respectively. We then exchange queries between branches:
\begin{equation}
\footnotesize
F_{f}^{\beta} = \operatorname{softmax}\left(\frac{Q_{\alpha}K_{\beta}}{d_{k}}\right)V_{\beta}, \quad
F_{f}^{\alpha} = \operatorname{softmax}\left(\frac{Q_{\beta}K_{\alpha}}{d_{k}}\right)V_{\alpha},
\end{equation}
where $d_k$ is a scaling factor. The outputs $F_{f}^{\alpha}$ and $F_{f}^{\beta}$ are combined into $F_{f}^{\gamma}$, which is then enhanced through cross-attention for feature intercalation:
\begin{equation}
\footnotesize
\{Q_{\gamma}, K_{\gamma}, V_{\gamma}\} = \mathcal{F}^{\gamma}_{qkv}(F_{f}^{\gamma}), \quad
\hat{F}_{f}^{\gamma} = \operatorname{softmax}\left(\frac{Q_{\gamma}K_{\gamma}}{d_{k}}\right)V_{\gamma}.
\end{equation}

Our dual-branch architecture effectively integrates text-prompt guidance across spatial and spectral dimensions. By embedding plain-residual-in-residual modules \cite{swinir,DRCT,rtcs} within a U-shaped cascade decoder \cite{restormer,Xrestormer}, our framework achieves both fine-grained and global restoration, effectively addressing various HSI degradations.

\input{tables/001_main_peersmethod_comparison}

\begin{figure*}
	\centering
	\includegraphics[width=0.98\textwidth]{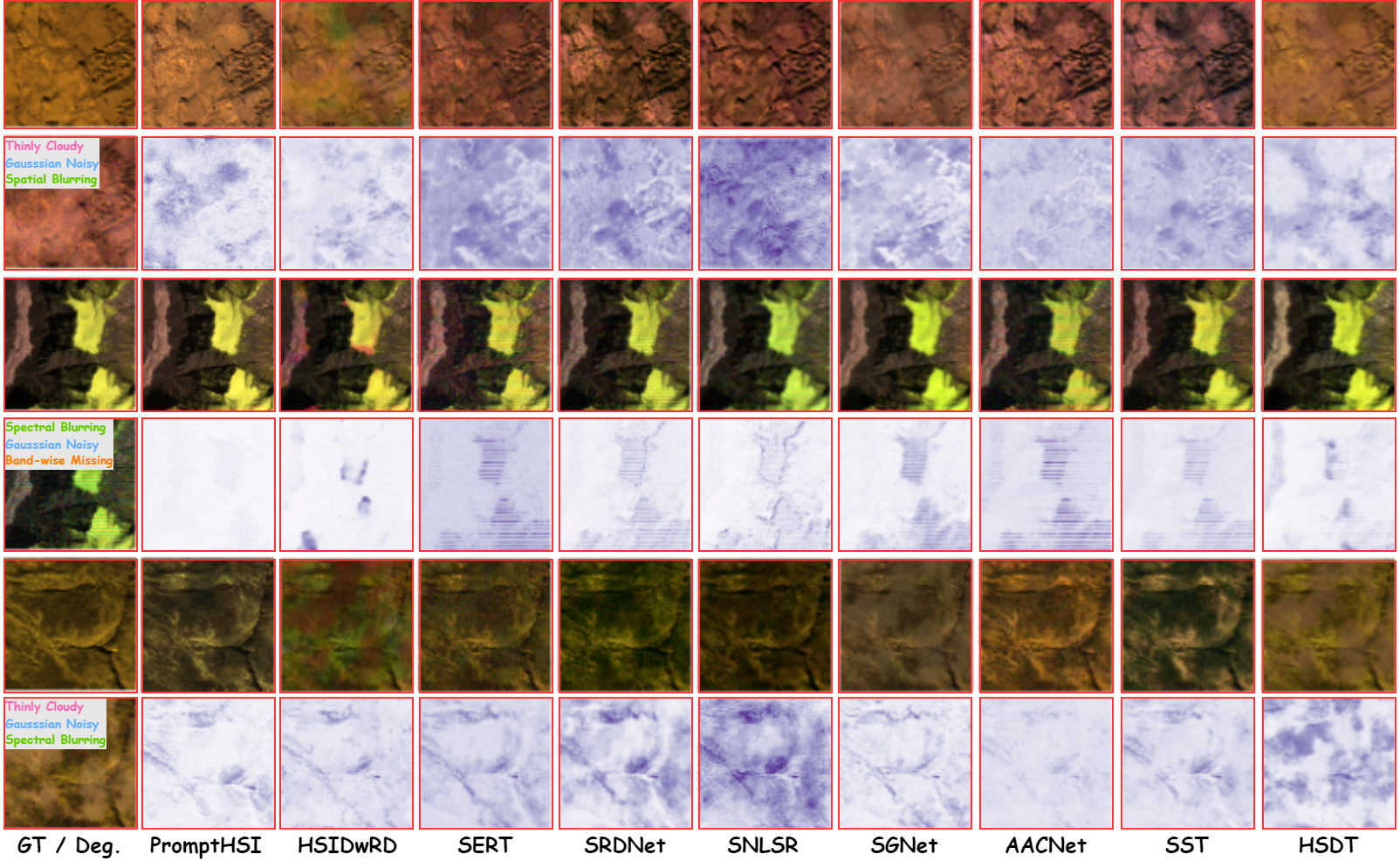}
\vspace*{-2mm}
	\caption{\small \underline{\textbf{Visualization results for HSI restoration methods.}}  \textbf{\emph{top}}: the reconstructed HSI, \textbf{\emph{bottom}}: the residual image derived from Ground-truth and reconstructed result. Most of state-of-the-arts HSI restoration methods fail to deal with composite degradations.
 }
	\label{fig:visual}
    \vspace{-2mm}
\end{figure*}

\subsection{Prompt-Guided Feature
Modulation}
\label{sec:4.2}




Directly using text features from VLMs for HSI restoration faces several critical challenges. Firstly, VLM features are trained on RGB-text pairs, lacking understanding of spectral characteristics and capturing complex spectral degradations. Secondly, their statistical distribution are mismatched because VLM features follow RGB image statistics, but HSI has essentially different data distributions, and direct adaptation requires complex domain translation. 

Instead of directly bridging the HSI-text domain gap, our proposed PGFM addresses the domain gap by  decomposing the problem into two easier sub-tasks. \ding{182} frequency-pattern matching with task descriptor and \ding{183} distribution adjustment by novel frequency-aware feature modulation mechanism. This strategy aim to selectively enhance spectral fidelity across frequency bands, thereby overcoming the limitations of text-guidance for avoiding direct cross-domain alignment between text features extracted from VLMs pre-trained on RGB-text domains and HSI.

\textbf{Task descriptor generation.} 
Given a text prompt $T_{text}$ that provides the corresponding degradation description to guide the HSI restoration network to obtain the specified restoration result. In this study, we employ CLIP \cite{CLIP}, which integrates vision and language information and performs effectively in text feature extraction. We freeze the pre-trained CLIP text encoder $\mathcal{E}^{CLIP}_{text}$ to maintain great linguistic consistency and add additional MLP layer $\Phi_{t}^{adpt.}$ as the feature adapter for pattern matching. With $\{\cdot\}_{e}$ denoting the frozen weights, this process can be expressed as:
\begin{equation}
\label{eq4}
\footnotesize
F_{text}=\Phi_{t}^{adpt.}(\{\mathcal{E}^{CLIP}_{text}\}_{e}(T_{text})),
\end{equation}
where $F_{text}$ denotes the task descriptors. With the task descriptors available, the next challenge is how to effectively incorporate this guidance into the HSI restoration process.

\textbf{Frequency-aware feature modulation}. Based on the frequency domain analysis, as illustrated in Section \ref{sec:deg}, we design our feature modulation mechanism to mimic how degradations affect different frequency components. Given an input feature map $F_{f}$, we first decompose it into frequency components:
\begin{equation}
\label{eq6}
\footnotesize
F_{f} = F_{f}^{l} + F_{f}^{h},
\end{equation}
where $F_{f}^{l}$ and $F_{f}^{h}$ represent low and high frequency components, obtained through frequency-domain filtering. 

Afterward, to merge text and image features, previous work \cite{textif} proposed employing a single MLP layer, $\Psi_{m}^{i}$, to integrate them with intensity and bias embeddings. Consequently, our feature modulation can be represented as:
\begin{equation}
\label{eq5}
\footnotesize
 \lambda_{m} = \Psi_{m}^{\uppercase\expandafter{\romannumeral1}}(F_{text}), \ \mu_{m} = \Psi_{m}^{\uppercase\expandafter{\romannumeral2}}(F_{text}),
\end{equation}
where $\Psi_{m}^{\uppercase\expandafter{\romannumeral1}}$ and $\Psi_{m}^{\uppercase\expandafter{\romannumeral2}}$ are the chunk operations of $\Psi_{m}$ to form the intensity and bias controller embeddings. To address the spectral consistency challenge in HSIs, our frequency-aware feature modulation adopts an affine transformation model (Eq.~\ref{eq2}) tailored for spectral data, formulated as:
\begin{equation}
\label{eq6}
\footnotesize
\hat{F}_{f} = (1 + \lambda_{m}^{l}) \odot F_{f}^{l} + (1 + \lambda_{m}^{h}) \odot F_{f}^{h} + \mu,    
\end{equation}
where Intensity controller $\lambda$ $\in$ [$\lambda_{m}^{l}$, $\lambda_{m}^{h}$] modulates the amplitude of high-frequency components to address noise, while Bias controller $\mu$ shifts the low-frequency baseline to compensate for blurring and band-missing effects. This design ensures frequency-specific adjustments that RGB AiO frameworks lack. 

To validate the efficacy of this mechanism, we performed ablation studies demonstrating improved spectral fidelity and PSNR consistency across spectral bands, underscoring the need for spectral domain modulation to manage complex HSI degradations effectively.


\vspace*{-1mm}
\subsection{Loss Function}
\vspace*{-1mm}
Given the wide spectral range in HSIs, using only the $\ell_1$ loss during training may not fully capture band variability; it tends to prioritize bands with larger values, potentially compromising spectral integrity.

\vspace*{-1mm}
\textbf{Spectral Angle Mapper Loss.} The SAM metric is essential for material identification and is less sensitive to intensity variations. It emphasizes angular discrepancies over magnitudes:
\begin{equation}
\footnotesize
\ell_{\text{SAM}}=\frac{1}{N} \sum_{n=1}^{N} \cos^{-1}\left(\frac{I_n^T I^*_n + \epsilon}{\|I_n\|_2 \cdot \|I_n^*\|_2 + \epsilon}\right),
\end{equation}
where $I_n$ and $I^*_n$ denote the $n$-th spectral vectors of the ground-truth and reconstructed HSIs, respectively, and $\epsilon$ prevents divergence.

\textbf{Stationary Wavelet Transform Loss.} SWT loss enforces frequency-domain consistency via wavelet sub-band alignment:
\begin{equation}  \label{eq:swt_fidelity}
\footnotesize
\ell_\text{SWT} =  \sum_{j} \lambda_{j} \big\| \text{SWT}(I)_j - \text{SWT}(I^*)_j \big\|_1,
\end{equation}
with $\lambda_{j}$ weighting the contribution of each sub-band.

\textbf{Band-wise MSE Loss.} BMSE loss balances reconstruction across all spectral bands to stabilize optimization:
\begin{equation}
\footnotesize
\ell_{\text{BMSE}} = \frac{1}{N} \sum_{n=1}^{N} \|I_n - I_n^*\|_2.
\end{equation}

\textbf{Total Loss.} The overall loss for training PromptHSI is defined as $\ell_\text{total} = \lambda_{1}\ell_\text{1} + \lambda_{2} \ell_\text{SAM} + \lambda_{3} \ell_\text{SWT} + \lambda_{4} \ell_\text{BMSE}$, where the coefficients are set to $\lambda_1=1$, $\lambda_2=0.001$, $\lambda_3=0.01$, and $\lambda_4=0.01$ throughout all experiments.

%% file: tables/001_main_peersmethod_comparison.tex
\begin{table*}[htbp]
\centering
\caption{\small \underline{\textbf{Overall performance evaluation for composite degradation and complexity comparison.}} To simulate composite degradations, we maintain a fixed $p$ of 0.5 for triggering individual degradation types, as demonstrated in Figure \ref{fig:second}, for all experiments. Note that $\textbf{Cr.}$ means $\textbf{Controllable}$ or $\textbf{Not}$. As for the complexity parts, M and G indicate $10^6$ and $10^9$, respectively. The best, second-best and third-best results are highlighted in $\best{bold-red}$, $\secondBest{bold-blue}$, and $\thirdBest{bold-green}$, respectively.}
\label{tab:performance}

\definecolor{input_color}{RGB}{245,245,245}      
\definecolor{one_to_one_color}{RGB}{240,248,255}  
\definecolor{specific_color}{RGB}{255,250,240}    
\definecolor{composite_color}{RGB}{245,255,250}   
\scalebox{0.77}{
\begin{tabular}{c|c|c|c|c|c|cccc|cc}
\toprule[0.15em]
\rowcolor{background_color}
\multicolumn{1}{c|}{Type} & Methods              & Venue $\&$ Year & Task       & Domain & \multicolumn{1}{c|}{Cr.} & \multicolumn{1}{c}{PSNR$\uparrow$}&\multicolumn{1}{c}{SAM$\downarrow$} & \multicolumn{1}{c}{RMSE$\downarrow$} & \multicolumn{1}{c|}{ERGAS$\downarrow$} & \multicolumn{1}{c}{$\#$Params$\downarrow$} & \multicolumn{1}{c}{FLOPs$\downarrow$} \\ \rowcolor{input_color}
\hline
Input&-&-&-&-& $\times$ & 15.2454 & 17.7463 & 0.0643 & 33.7220 & - & - \\\rowcolor{one_to_one_color}
\hline
&
HSIDwRD \cite{HSIDwRD} & ICCV 2021 & Denoising  & HSI & $\times$ & 20.6270 & 12.7891 & 0.0364 & 97.3601 & 23.62M & 5054.16G \\\rowcolor{one_to_one_color}& 
SERT \cite{SERT} & CVPR 2023 & Denoising & HSI & $\times$ & 16.1657 & 15.9853 & 0.0348 & 29.2053 & 2.74M & 176.06 G \\\rowcolor{one_to_one_color}& 
SRDNet \cite{SRDNet} & TGRS 2024 & Super-Res. & HSI & $\times$ & 19.0520 & 11.9590 & 0.0320 & \thirdBest{16.5861} & 1.86M & 3771.80G \\\rowcolor{one_to_one_color}&
SNLSR \cite{SNLSR} & TIP 2024 & Super-Res. & HSI & $\times$ & 17.8785 & 20.4389 & 0.0445 & 19.1267 & 1.75M & 14.20G \\\rowcolor{one_to_one_color}
One-to-One& 
SGNet \cite{SGNET} & ISPRS 2022 & Decloud & HSI & $\times$ & 18.0694 & 12.9476 & 0.0359 & 17.9773 & 0.54M & 57.39G \\\rowcolor{one_to_one_color}& 
AACNet \cite{AACNet} & TGRS 2023 & Decloud & HSI & $\times$ & 17.6279 & 14.5962 & 0.0324 & 26.2115 & 2.76M & 171.23G \\\rowcolor{one_to_one_color}&
SST \cite{SST} & AAAI 2023 & Inpainting & HSI & $\times$ & 18.9792 & 10.9860 & \thirdBest{0.0244} & 26.9934 & 4.30M & 281.72G \\\rowcolor{one_to_one_color}&
HSDT \cite{HSDT}& ICCV 2023 & Inpainting & HSI & $\times$ & 18.0665 & 11.3451 & 0.0372 & 68.4451 & 0.51M & 67.55G \\\rowcolor{specific_color}
 \hline
 & 
AirNet \cite{AirNet} & CVPR 2022 & - & RGB & $\times$ & \secondBest{23.6647} & \secondBest{7.5970} & \secondBest{0.0233} & \secondBest{16.4194} & 6.07M & 245.86G \\\rowcolor{specific_color}

{{All-in-One}}& 
PromptIR \cite{PromptIR} & NeurIPS 2023 & - & RGB & $\times$ & 17.2667 & \thirdBest{9.6437} & 0.0405 & 29.2768 & 33.18M & 132.07G \\\rowcolor{specific_color} 
for specific&
HAIR \cite{HAIR}
& arXiv 2024 & - & RGB & $\times$ & \thirdBest{20.5779} & 14.4329 & 0.0275 & 24.4211 & 7.86M & 42.51G \\\rowcolor{specific_color}
 &
InstructIR \cite{InstructIR} & ECCV 2024 & - & RGB & $\checkmark$ & 20.3106 & 14.5727 & 0.0306 & 27.5454 & 16.02M & 17.28G \\ \hline
\rowcolor{composite_color}
All-in-One& OneRestore \cite{OneRestore} & ECCV 2024 & - & RGB & $\checkmark$ & 17.3321 & 15.0320 & 0.0271 & 52.1162 & 5.85M & 12.05G \\ \rowcolor{composite_color}
for composite & \textbf{PromptHSI (Ours)} & - & - & HSI & $\checkmark$ & \best{26.4432} & \best{6.0601} & \best{0.0185} & \best{8.3566} & 26.14M & 135.83G \\ 
\bottomrule[0.15em]
\end{tabular}}
\end{table*}

%% file: sections/experiments.tex
\section{Experimental Results}
\label{sec:expALL}

\input{tables/002_degradation}
\input{tables/003_ablationstudy}


\textbf{Experiment Setup.} This study employed the AVIRIS dataset provided by Lin \emph{et al.} \cite{QRCODE,CSAKD}, an extensive HSI dataset with numerous bands spanning diverse geographical areas, for experiments. The dataset comprises 2,078 HSIs, which were randomly divided into training (1,678 images), validation (200 images), and testing (200 images) sets. As illustrated in the \textbf{Supplementary}, all clean HSIs were pre-processed to synthesize composite degraded HSIs along with corresponding text descriptions. This dataset is publicly available to facilitate further research.

\subsection{Quantitative Results and Analysis}

To evaluate the performance of PromptHSI, we compared it with eight other supervised HSI restoration methods for specific degradation, including HSIDwRD \cite{HSIDwRD}, SERT \cite{SERT}, SRDNet \cite{SRDNet}, SNLSR \cite{SNLSR}, SGNet \cite{SGNET}, AACNet \cite{AACNet}, SST \cite{SST}, and HSDT \cite{HSDT} in the synthesized AVIRIS dataset with composite degradations. Because there is no AiO HSI restoration method, we compare four AiO RGB image restoration methods in the experiments: AirNet \cite{AirNet}, PromptIR \cite{PromptIR}, HAIR \cite{HAIR}, InstructIR \cite{InstructIR}, AdaIR \cite{adair}, and OneRestore \cite{OneRestore}. The model performance was measured using the four metrics, including PSNR, SAM, RMSE, and ERGAS. Table \ref{tab:performance} presents the quantitative results. 


\textbf{Ours vs. Task-specific HSI restoration methods.} All peer-methods fail to effectively recover the degraded HSIs due to their ability to learn different or unseen degradations. In contrast, PromptHSI can restore better even without text-prompt, as shown in Table \ref{table:ablation_component}, which exhibits the capability of our unified restoration network design.

\textbf{Ours vs. All-in-One RGB restoration methods.} Unlike AiO RGB image restoration methods that rely on uniform feature modulation, our PromptHSI framework’s frequency-aware modulation dynamically tailors restoration to each spectral band, thereby achieving cross-band consistency crucial for HSI applications.

In general, our PromptHSI exceeds other AiO methods. Firstly, most of existing methods struggle to handle composite degradations. As discussed previously, both data and features extracted by pre-trained CLIP \cite{CLIP} exhibit domain gaps with HSI. Therefore, these models are not effective in transferring to AiO HSI restoration. 

Specifically, OneRestore \cite{OneRestore} integrated task description features into its encoder, which hindered its ability to capture spatial-spectral correlations in HSIs due to a domain gap, leading to decoding failures.




\subsection{Ablation Studies and Analysis}
\label{sec:ablation_studies}

\begin{figure}
	\centering
	\includegraphics[width=0.455\textwidth]{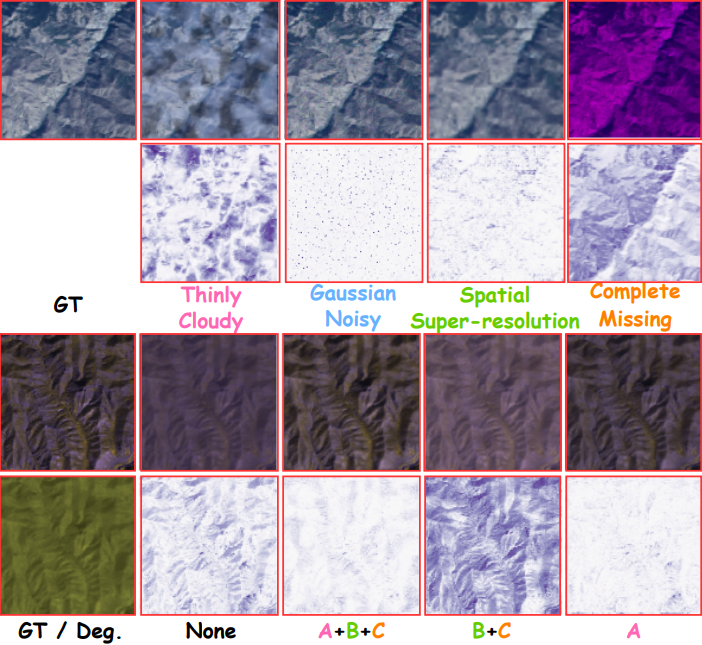}
	\caption{\small \underline{\textbf{Visualization results for the controllability.}}  \textbf{\emph{Top}}, for single-specific degradation, the plots in the first row represent the degraded HSI, while the second row shows residual maps, obtained from the reconstructed HSI, with or without the guidance provided by the text-prompt. \textbf{\emph{Bottom}}, in the case of composite degradation, the top row of plots depicts the recovered HSI, while the residual maps are obtained from the reconstructed HSI using the specified text-prompt guidance with the ground-truth.}
	\label{fig:control}
    \vspace{-2mm}
\end{figure}

The aforementioned experimental results demonstrate the effectiveness of our proposed PromptHSI framework under composite degradations. We further conduct ablation studies to investigate the impact of different components and strategies.
%

\textbf{Integration strategy of frequency-aware modulation.}
As shown in Table \ref{table:ablation_condition_information}, modulating frequency via prompting
demonstrates superior effectiveness compared to the weighted-sum manner, where we implemented an MLP layer for dynamic weighted averaging of inputs and outputs. The intensity-only configuration outperforms the bias-only, with a notably larger advantage in the ERGAS metric, suggesting that this integration strategy effectively guides network restoration across most scenarios.


\textbf{Visual-prompt or text-prompt.} The type of prompting is crucial for recognizing degradations and guiding model behavior. In our first configuration, we use RGB bands from HSIs as visual prompts, encoded via CLIP \cite{CLIP} and guided by PGFM. In the second, we combine RGB and textual prompts to form hybrid task descriptors. In the final configuration, PGFM is removed and we employ the PIM and PGM blocks from PromptIR \cite{PromptIR} for generating visual prompts and feature interaction.

Table \ref{table:prompt} shows that composite spatial degradations cause significant information loss, making it difficult to obtain compact task descriptors or effective visual prompts. Incorporating textual features overcomes these challenges by providing robust semantic guidance.




\textbf{Ablation of network component.} To understand the impact of the network components, we performed ablation studies by removing PGFM, adapter and spatial spectral enhancement from the network, with results shown in Table \ref{table:ablation_component}. Even without textual features, our proposed PromptHSI outperforms other peer methods in handling composite degradation, demonstrating the effectiveness and universality of our network architecture. The adapter proves crucial in bridging the domain gap between the VLM and HSI features via pattern matching, leading to improved global reconstruction performance, particularly evident in the ERGAS metric. Lastly, we can observe that there is a drop with or without spatial-spectral enhancement, which represents that feature enhancement is essential for details recovery within cascade decoder.

These results demonstrate PromptHSI's superior capability in maintaining spectral fidelity across multiple degradation types, highlighting the necessity of frequency-aware modulation, particularly when compared to AiO RGB methods that exhibit substantial performance degradation in the HSI domain. In summary, PromptHSI offers a universal and controllable solution for addressing composite degradation in real-world applications.


%% file: tables/002_degradation.tex
\begin{figure}
	\centering
\includegraphics[width=0.48\textwidth]{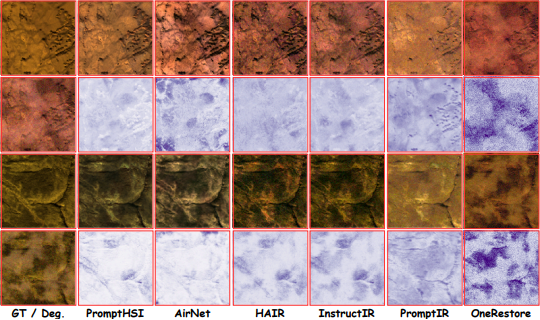}
	\caption{
 \small \underline{\textbf
 {
 Comparison for AiO RGB image restoration methods.}} For the text-prompt-based methods \cite{InstructIR,OneRestore} and our PromptHSI, we employ the standard text-prompts presented in this study for training network and inference for related experiments.
 }
	\label{fig:vis-all-in-one}
\end{figure}

\begin{figure}
		\centering
		\includegraphics[width=0.48\textwidth]{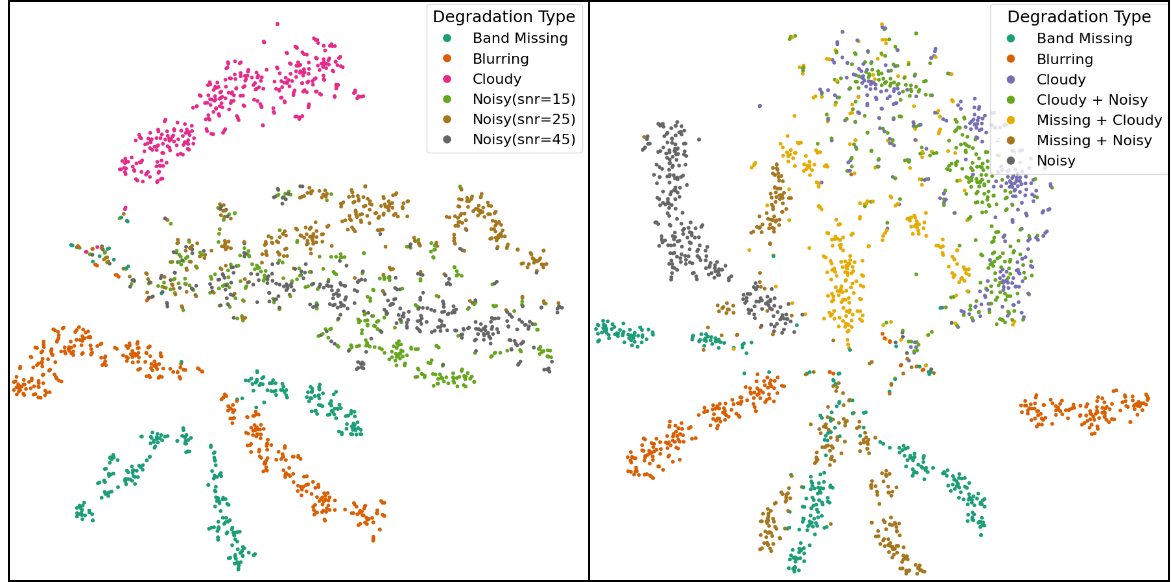}
		\caption{\small \underline{\textbf{Comparison of t-SNE visualization results}} for the degradation embeddings of the proposed PromptHSI trained on single degradation. (a) for multiple degradation, (b) for composite degradatation. The proposed PromptHSI have demonstrated its discriminative capability in identifying different degradation even unseen degradation and its combination during testing-time.}
		\label{fig:tsne}	
\end{figure}

%% file: tables/003_ablationstudy.tex
\begin{table*}[t]
\parbox{.33\linewidth}{
\centering
\caption{\small Ablation of integration strategy. }
\label{table:ablation_condition_information}

\definecolor{best_color}{RGB}{230,255,230}  
\definecolor{other_color}{RGB}{255,230,230}  
\vspace{-3mm}
\setlength{\tabcolsep}{2pt}
\scalebox{0.69}{
\begin{tabular}{c | c }
\toprule[0.15em]
\rowcolor{background_color}
Strategy & PSNR / SAM / RMSE / EGRAS \\\rowcolor{other_color}
\hline
Intensity-only & \secondBest{26.0070} / \secondBest{6.3636} / \secondBest{0.0187} / \secondBest{9.7608} \\\rowcolor{other_color}
Bias-only & 25.6384 / 6.3963 / 0.0190 / 12.4992 \\\rowcolor{other_color}
Weighted-sum & 25.7284 / 6.3561 / 0.0188 / 14.3605 \\ \rowcolor{best_color}\hline
\textbf{Intensity \& Bias}& \best{26.4432} / \best{6.0601} / \best{0.0185} / \best{8.3566} \\
\bottomrule[0.15em]
\end{tabular}}}
\hfill
\parbox{.30\linewidth}{
\centering
\caption{\small Effect of prompting method. }
\label{table:prompt}
\definecolor{best_color}{RGB}{230,255,230}  
\definecolor{other_color}{RGB}{255,230,230}  
\vspace{-3mm}
\setlength{\tabcolsep}{2pt}
\scalebox{0.69}{
\begin{tabular}{c | c  | c }
\toprule[0.15em]
\rowcolor{background_color}
Prompt type & Cr. & PSNR / SAM / RMSE / EGRAS \\\rowcolor{other_color}
\hline
RGB to VLM & $\times$ & 24.1273 / 7.6236 / 0.0257 / 14.5208 \\\rowcolor{other_color}
Hybrid to VLM & $\checkmark$ & \secondBest{25.2385} / \secondBest{6.8974} / \secondBest{0.0191} / \secondBest{11.4359}
\\\rowcolor{other_color}
PGB \& PIM \cite{PromptIR} &  $\times$ &  23.0567 / 9.4862 / 0.0287 / 24.5575\\ \rowcolor{best_color}\hline
\textbf{Text to VLM}& $\checkmark$ & \best{26.4432} / \best{6.0601} / \best{0.0185} / \best{8.3566} \\
\bottomrule[0.15em]
\end{tabular}}}
\parbox{.36\linewidth}{
\centering
\caption{\small Ablation of network component.} 
\label{table:ablation_component}
\definecolor{best_color}{RGB}{230,255,230}  
\definecolor{other_color}{RGB}{255,230,230}  
\vspace{-3mm}
\setlength{\tabcolsep}{2pt}
\scalebox{0.69}{
\begin{tabular}{c | c }
\toprule[0.15em]
\rowcolor{background_color}
Without & PSNR / SAM / RMSE / EGRAS \\\rowcolor{best_color}
\hline
\textbf{Full model} & \best{26.4432} / \best{6.0601} / \best{0.0185} / \best{8.3566} \\ \rowcolor{other_color}\hline
PGFM & 25.4905 / 6.6348 / 0.0205 / 13.2072 \\\rowcolor{other_color}
adapter & 25.8686 / {6.3561} / {0.0188} / 14.3605 \\\rowcolor{other_color}
enhancement & 23.7511 / 7.9918 / 0.0262 / 16.0007 \\
\bottomrule[0.15em]
\end{tabular}
}}
\hfill
\vspace*{-1mm}
\end{table*}

%% file: sections/conclusion.tex
\section{Conclusion}
\label{sec:conclusion}
In this paper, we introduced PromptHSI, the first universal framework for hyperspectral image restoration that effectively addresses composite degradations—a challenge that traditional methods have struggled to overcome. Unlike existing AiO RGB restoration approaches, which are hampered by domain gaps and inadequately capture HSI-specific degradation patterns, our method is specifically tailored to the unique attributes of HSIs. By integrating frequency-aware modulation with vision-language model–guided prompt learning, PromptHSI bridges textual cues and frequency-domain information, enabling adaptive and controllable restoration. Our approach significantly outperforms state-of-the-art methods across various degradation scenarios. Moreover, we present and release the first composite degradation dataset for HSI restoration, paving the way for future research in this field.

%% file: sections/overview_sup.tex
\section{Appendix}
\noindent \textbf{Overview.} This supplementary material provides a more detailed explanation of the model configuration, the preparation of the dataset, and additional results of the experiments. \ding{182} First, we provide a detailed experiment setup, dataset description, and its preparation pipeline in Sections \ref{sec:supexp} and \ref{sec:supdata}. \ding{183} Secondly, we provide the illustration of the PromptHSI decoder module and present our model configuration in Section \ref{sec:suparch}. \ding{184} Next, we present additional experiment results and visualization in Section \ref{sec:supexperiments}.

\section{Experiment Setup}
\label{sec:supexp}

\noindent \textbf{Dataset description.} This study used a dataset captured by the Airborne Visible/Infrared Imaging Spectrometer \footnote{AVIRIS Data Portal: \hyperlink{https://aviris.jpl.nasa.gov/data/index.html}{https://aviris.jpl.nasa.gov/data/index.html}}(AVIRIS) \cite{35,QRCODE}. The \footnote{HyperAD: \hyperlink{https://github.com/chingheng0808/HyperAD}{https://github.com/chingheng0808/HyperAD}}dataset encompasses a wide range of natural settings throughout the United States and Canada, including urban areas, mountains, lakes, agricultural regions, and diverse vegetation, gathered from 2006 to 2011. The original HSIs were partitioned into distinct subimages, each measuring 259 $\times$ 259 pixels, composed of 224 spectral bands spanning wavelengths from 400 to 2500 nm. Consistent with prior research \cite{ADMMADAM,QRCODE,CSAKD}, spectral bands identified as lower quality (1-10, 104-116, 152-170, and 215-224) were excluded, resulting in HSIs with 172 spectral bands.

\noindent \textbf{Hyperparameter setting.} We used single NVIDIA-A100 for training. The batch size was set to 8, and the number of training epochs was fixed at 300 for all experiments involving the proposed method. The AdamW optimizer \cite{AdamW} ($\beta_1=0.9$, $\beta_2=0.999$, weight decay $1 \times 10^{-2}$) was employed, with an initial learning rate of $3 \times 10^{-4}$. The learning rate was dynamically adjusted during the training process using a step-based learning rate scheduler. The scheduler reduces the learning rate by a factor of 0.5 every 50 epochs to ensure stable convergence. 

\noindent \textbf{Data augmentation.} The input HSI was randomly cropped to patches with the size of $112 \times 112 \times 172$, incorporating random horizontal, vertical rotations and flipping during the training phase. During inference, the model was evaluated using input dimensions of $224 \times 224 \times 172$.

\begin{table}[ht]
\centering
\caption{\small \underline{\textbf{Inference time comparison.}} For our simulation, we use randomly generated tensor with the size of $224 \times 224 \times 172$, conducting inference $30$ times and taking the average for evaluation on All-in-One RGR image restoration methods.}
\label{tab:infer_time}
\scalebox{0.82}{
\begin{tabular}{r|c}
\toprule[0.15em]
\multirow{1}{*}{\textbf{Method}} & \multicolumn{1}{c}{Inference time per inputs ↓ ($\mu$ ± $\sigma$)} \\\rowcolor{cnn_color}\hline
\textbf{AirNet} \cite{AirNet}& 230.54 ± 59.61 ms\\\rowcolor{cnn_color}
\textbf{PromptIR} \cite{PromptIR} & 178.58 ± 47.92 ms  \\\rowcolor{cnn_color}
\textbf{HAIR} \cite{HAIR} & 162.33 ± 38.88 ms\\\rowcolor{cnn_color}
\textbf{AdaIR} \cite{adair} & 107.93 ± 59.86 ms\\\rowcolor{cnn_color}
\textbf{InstructIR} \cite{InstructIR} & 32.62 ± 8.63 ms\\\rowcolor{cnn_color}
\textbf{OneRestore} \cite{OneRestore} & 24.90 ± 5.94 ms\\\rowcolor{ours_color}\hline
\textbf{PromptHSI (Ours)}& 282.78 ± 74.48 ms  \\

\bottomrule[0.15em]
\end{tabular}}
\end{table}

%% file: sections/dataset_sup.tex
\section {Dataset Preparation}
\label{sec:supdata}

\begin{figure*}[htbp]
		\centering
		\includegraphics[width=1.0\textwidth]{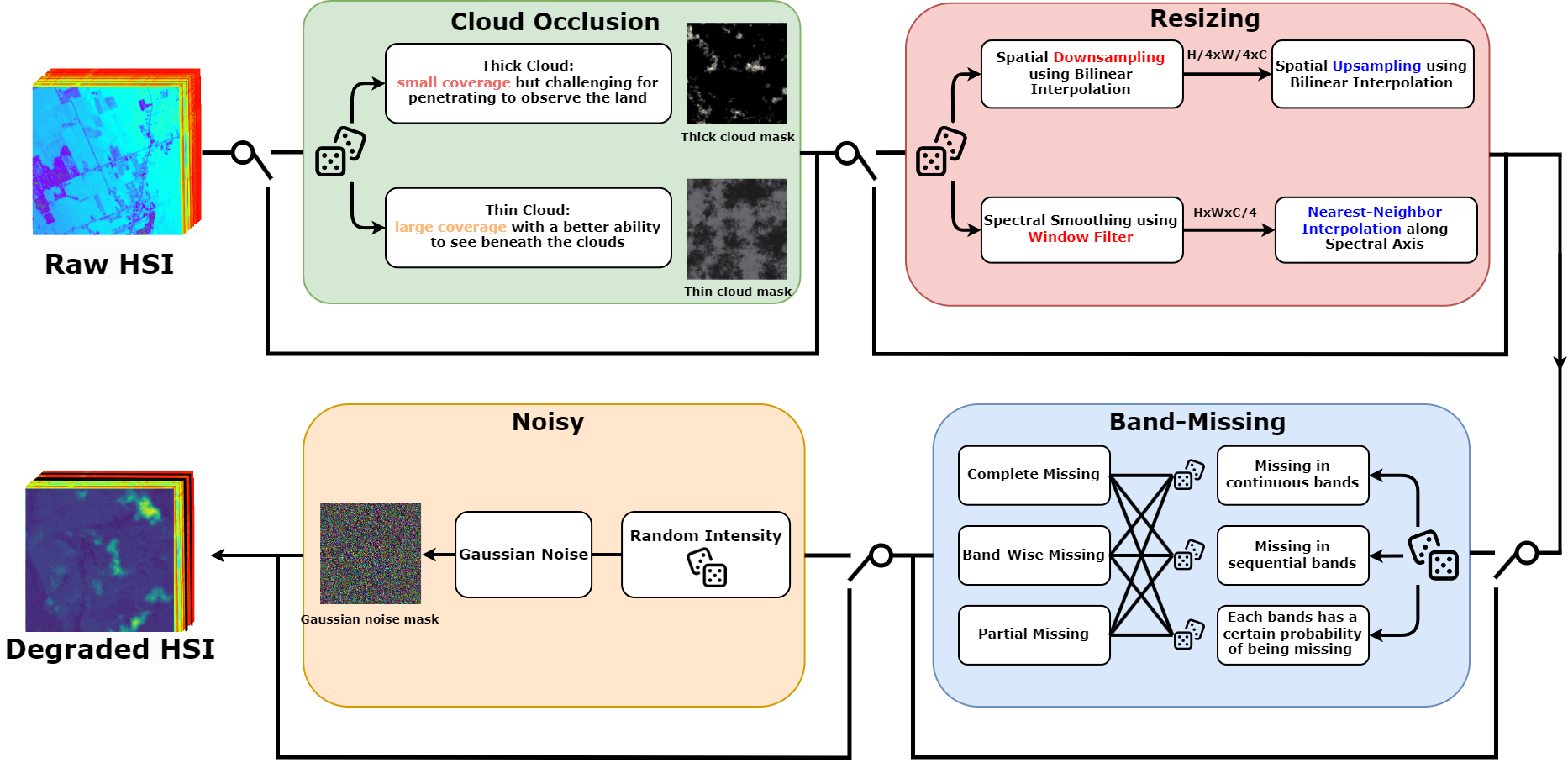}
		\caption{\small \underline{\textbf{The proposed dataset synthesis pipeline.}} Each degradation is activated with probability $p$. Note that we set $p$ = $0.5$ for all experiments. For the diversity of synthetic dataset, each active degradation chooses and implements one of its subdegradations on the input HSI.}
		\label{fig:dataset}	
\end{figure*}

To the best of our knowledge, there have been no reliable simulations and analysis of composite degradation in HSI restoration. Therefore, designing a realistic degradation process that accurately simulates real-world scenarios is important. To address this, we modify the GDM (Gated Degradation Model) \cite{GDM} from the RGB image blind super-resolution task to accommodate our task.

\noindent \subsection{Degradation Model} 

It is essential to consider the complete imaging acquisition pipeline when simulating the hyperspectral imagery process. Specifically, the observed images are initially affected by cloud coverage or shadow occlusion, resulting in ground surface invisibility. Subsequently, the image resolution is altered due to hardware limitations. The imaging components then introduce a random noise into the image. Finally, at the transmission, atmospheric interference or bandwidth constraints lead to the manifestation of stripe noise.

\noindent \textbf{Classical hyperspectral imaging model.} HSI restoration is an ill-posed problem, which assumes the clean image is affected by multiple types of degradation. Mathematically, the degraded image $I_{\text{degraded}}$ is generated from the clean image $I_{\text{clean}}$ as follows:
\begin{equation}
\footnotesize
I_{\text{degraded}} = D_{stripe,n,s,c}(I_{\text{clean}}) = \left[\downarrow (c  \oplus I_{\text{clean}})  + n\right] \odot M_{stripe},
\end{equation}
First, clean HSI $I_{\text{clean}}$ is degraded by cloud occlusion (denoted by $c$). Then, the occluded HSI is down-sampled (denoted as $\downarrow(\cdot)$) and an additive white Gaussian noise (denoted as $n$) is added to the degraded HSI. Finally, some of the HSI $I_{\text{degraded}}$ bands are corrupted and missed by the stripe effect during data transmission (denoted $stripe$).

\noindent \textbf{Cloud occlusion degradations.} We followed the Satellite Cloud Generator \cite{cloudgenerator} to synthesize the cloud and degradation. The cloudy HSI $I_{\text{cloudy}}$ is generated by blending the clear HSI $I_{\text{clean}}$ with the cloud HSI $I_{\text{cloud}}$ using the cloud transparency mask $M_C$:
\begin{equation}
\footnotesize
I_{\text{cloudy}} = (1 - M_C) \cdot I_{\text{clean}} \oslash  M_C \cdot I_{\text{cloud}},
\end{equation}
where $M_C$ represents the transparency mask that controls the influence of the cloud, $I_{\text{cloud}}$ is the cloud component image, $\oslash $ denotes the blending operator.

\noindent \textbf{Noise and stripe effect degradations.} Noise introduced during data acquisition can significantly affect image quality. Two common types of noise in HSI are Additive White Gaussian Noise (AWGN) and stripe noise. Stripe pattern noise appears as vertical or horizontal lines in specific bands, typically caused by sensor calibration problems and data transmission. When both AWGN and stripe pattern noise are present in spatial location $(i, j)$ in spectral band $k$, the noisy signal can be represented as:

\begin{equation}
\footnotesize
I_{\text{noisy}}(i, j, k) = \left[I_{\text{clean}}(i, j, k) + N_{\text{AWGN}}(i, j, k)\right] \odot M_S.
\end{equation}

\noindent \textbf{Probabilistic gated degradation model.} Following GDM \cite{GDM}, which aims to synthesize the dataset with composite degradations for RGB image, we redesign its degradation types to better suit the attribute of HSI. The gate mechanism is introduced to \textit{randomly} select the base degradation types to be included in the degradation process. Formally, the probabilistic gated degradation process $D_G$ can be expressed as: 
\begin{equation} 
\footnotesize 
I_{\text{degraded}} = D_G(I_{\text{clean}}) = (\sigma_g(D_1) \circ \sigma_g(D_2) \cdots \sigma_g(D_m))(I_{\text{clean}}), \end{equation} 
where $\sigma_g(D_i)(r^d) = \begin{cases} D_i(r^d), & g_i \sim \text{Bernoulli}(p), 
\\ r^d, & g_i = 0, \end{cases}$ and $r^d$ denotes the input HSI. In this case, each gate $g_i$ is activated with a probability $p$ (e.g., $p = 0.5$), which determines whether a specific degradation $D_i$ is applied. PGDM enables the diverse integration of fundamental degradation types, enhancing the synthesized HSI to be more vibrant in real-world scenarios.








\begin{table*}[htbp]
    \renewcommand{\arraystretch}{1.1}
    \centering
    \caption{\small \underline{\textbf{The details configurations and hyperparameter settings for degradation synthesis.}} The configuration for the Cloudy degradation is based on the parameters of Satellite Cloud Generator \cite{cloudgenerator}. As for $K$ in the band-missing configuration, it represents a value between 0 and 172 (the total number of HSI bands). The last column displays the PSNR and SAM metrics for the degraded HSIs, which contain only the corresponding degradation in the testing set.}
    \label{tab:degradation_config}
    \begin{adjustbox}{max width=\textwidth}
    \begin{tabular}{c|c|p{0.5\textwidth}|c} 
    \Xhline{2pt}
    Degradations & Types & Configurations & Mean PSNR / SAM on specific degradation\\
    \hline
    \multirow{8}{*}{Cloudy}  
    & \centering \multirow{4}{*}{Thickly} & 
    \texttt{locality\_degree=[2, 4],
    min\_lvl=0.0,
    max\_lvl=1.0,
    clear\_threshold=[0.0, 0.4],
    blur\_scaling=1,
    cloud\_color=True,
    channel\_offset=0,
    decay\_factor=1.0}
    & \multirow{4}{*}{11.0379 / 4.0143} \\
    \cline{2-4}
    & \multirow{4}{*}{Thinly} & 
    \texttt{locality\_degree=1,
    min\_lvl=[0.0, 0.4],
    max\_lvl=[0.4, 0.6],
    clear\_threshold=0.0,
    blur\_scaling=2,
    cloud\_color=True,
    channel\_offset=0,
    decay\_factor=1.0}
    & \multirow{4}{*}{7.0316 / 17.1393} \\
    \hline
    \hline
    \multirow{11}{*}{Blurring}  
    & \multirow{5}{*}{Spatial} & 
    \texttt{\textbf{Bilinear} interpolation is used to downsample by a factor of 4, followed by \textbf{bilinear} interpolation to restore the image to its original size in the spatial domain.}
    & \multirow{5}{*}{22.2780 / 3.8414} \\
    \cline{2-4}
    & \multirow{6}{*}{Spectral}  & 
    \texttt{Using a \textbf{Gaussian window filter} with a \textbf{window size of 5 and stride of 4} to smooth along the spectral dimension, followed by \textbf{nearest-neighbor} interpolation to restore the image to its original size in the spectral domain.}
    &  \multirow{6}{*}{24.8798 / 12.6105} \\
    \hline
    \hline
    \multirow{2}{*}{Noisy} & \multirow{2}{*}{Gaussian noise} & 
    \texttt{Using SNR to control the noise intensity, where $\text{SNR} \sim \mathcal{N}(\mu, \sigma)$, with $\mu = 35$ and $\sigma = 5$.}
    &  \multirow{2}{*}{16.2706 / 12.9214} \\
    \hline
    \hline
    \multirow{9}{*}{Band-missing}  
    & \multirow{2}{*}{Complete} & 
    \texttt{All pixel values in randomly chosen $K$ bands are set to 0.}
    & \multirow{2}{*}{28.2688 / 26.6282} \\
    \cline{2-4}
    & \multirow{4}{*}{Band-wise} & 
    \texttt{On randomly chosen $K$ bands, pixels in a predetermined sequence of rows (for instance, the 1st, 3rd, 5th, 7th, 9th, etc.) are set to 0.}
    &  \multirow{4}{*}{30.1465 / 5.5401} \\
    \cline{2-4}
    & \multirow{3}{*}{Partial} & 
    \texttt{On randomly chosen $K$ bands, each row in the bands has a 0.3 probability of being set to 0.}
    &  \multirow{3}{*}{27.8382 / 5.8436} \\
    \Xhline{2pt}
    \end{tabular}
    \end{adjustbox}
\end{table*}

\subsection{Degradation Simulation}

Following the exposition in the previous subsection, where we detailed the simulation methodology for hyperspectral image degradation, we now proceed to elaborate on the considerations and processes involved in simulating each specific type of degradation.

%

\subsubsection{Cloud Occlusion}
\noindent \textbf{Background.} Although optical satellite imagery can be applied to various tasks, it is often affected by weather conditions, such as cloud cover, making it difficult to clearly observe the appearance of surface features. As a result, compiling high-quality datasets that are affected by cloud cover proves to be a difficult task, making the generation of high-quality cloudy data necessary.

\begin{figure}
        \centering
		\includegraphics[width=0.4\textwidth]{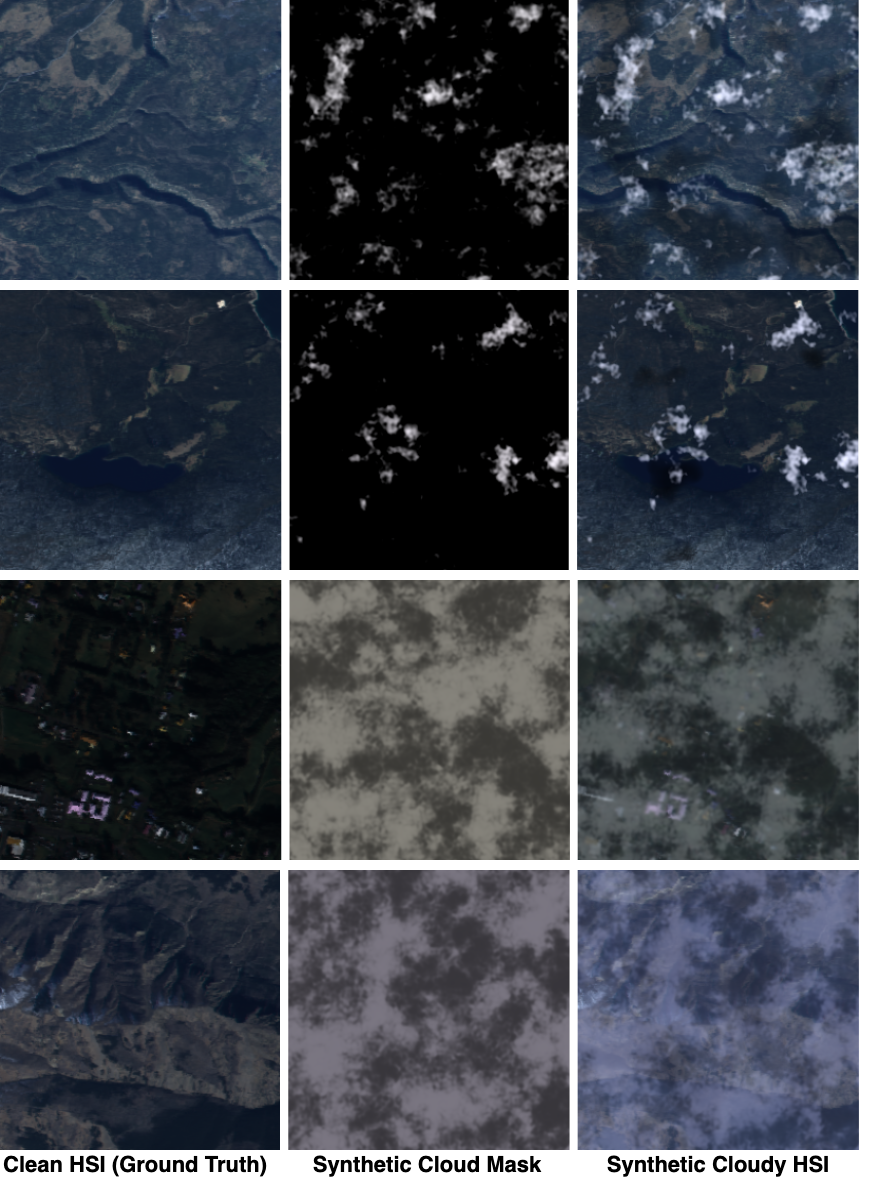}
		\caption{\small \underline{\textbf{Visualization of synthetic cloudy HSI.}} The first two rows illustrate thick cloud cover conditions, whereas the last two rows demonstrate thin cloud cover conditions. Thick clouds cover a smaller portion of the overall HSI, but the areas obscured by these clouds are difficult to observe. In contrast, thin clouds, while covering a larger area, allow for faint visibility of the underlying surface features, somewhat resembling a foggy effect.}
		\label{fig:cloudy}	
\end{figure}

\noindent \textbf{Synthesis.} Following the method for synthesizing clouds in \footnote{The official Github repository of Satellite Cloud Generator \cite{cloudgenerator}: \hyperlink{https://github.com/strath-ai/SatelliteCloudGenerator.git}{https://github.com/strath-ai/SatelliteCloudGenerator.git}} Satellite Cloud Generator \cite{cloudgenerator}, we first use a pair of real HSIs, one containing clouds and the other a neighboring image. 
Afterwards, using the cloud mask detector proposed in \cite{clouddetector}, we calculate the cloud mask area in the spectral bands of Sentinel-2. This mask is then used to approximate the cloud end-member function in different spectra, and finally, we generate clouds on the given clean HSI that contain thinly or thickly cloudy HSI, as shown in Figure \ref{fig:cloudy}.

\begin{figure}
		\centering
		\includegraphics[width=0.42\textwidth]{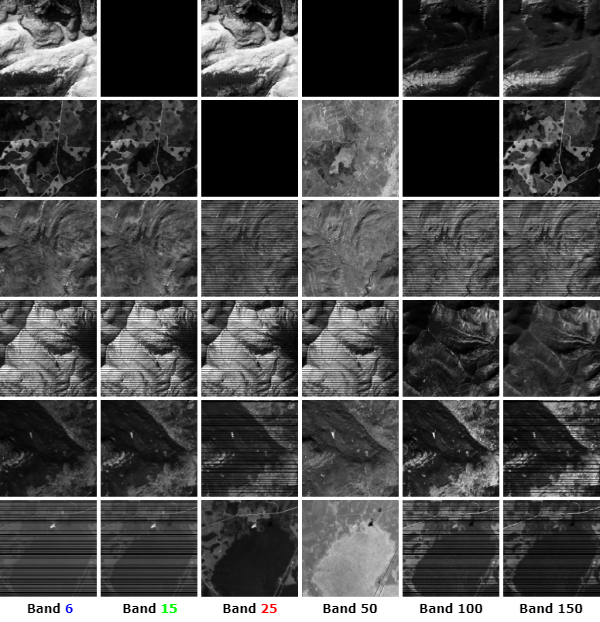}
		\caption{\small \underline{\textbf{Visualization of simulated band missing on HSI}}. We selected multiple bands to demonstrate different types of band missing degradation, with the 6th, 15th, and 25th bands representing the \textcolor[rgb]{0., 0., 1.}{B}, \textcolor[rgb]{0., 1., 0.}{G}, and \textcolor[rgb]{1., 0., 0.}{R} bands, respectively.The first two rows depict the
        band complete missing case, the next two rows indicate instances of band-wise missing, and the final two rows represent cases of partial missing.}
		\label{fig:bandmissing}	
\end{figure}

\subsubsection{Spatial Blurring and Spectral Blurring}
\noindent \textbf{Background.} 
Due to diverse sensor payloads and varying orbital altitudes of Low Earth Orbit (LEO) satellites, remote sensing data exhibits significant heterogeneity in both spatial and spectral resolutions. Hyperspectral imaging sensors, in particular, often suffer from limited spatial resolution due to technical and cost constraints, hindering detailed surface feature observation. To address these limitations, we incorporate both spatial super-resolution and spectral super-resolution tasks into our model framework. 

\noindent \textbf{Synthesis.} To simulate resolution degradation, we implement a downsampling-upsampling procedure across spatial or spectral domains. First, the formulations for spatial blurring process can be defined as follows:
\begin{equation}
\footnotesize
    I_{\text{spatial blurring}} = {I_{\text{clean}}}_{Spa_{\downarrow }^{4}Spa_{\uparrow }^{4}},
\end{equation}
where $Spa_{\downarrow }$ and $Spa_{\uparrow }$ denotes as spatial bilinear down-sampling and bilinear interpolation up-sampling with factor of $4$, respectively. Subsequently, the spectral blurring process can be expressed as:
\begin{equation}
\footnotesize
    I_{\text{spectral blurring}} = (\Phi(I_{\text{clean}}, 5, 4))_{Spe_{\uparrow }^{4}},
\end{equation}
where $\Phi(\cdot, \omega, \varsigma)$ denotes a sliding Gaussian window filter with window size $\omega$ and stride $\varsigma$, respectively, to reduce the spectral resolution. $Spe_{\uparrow}$ represents nearest-neighbor interpolation with factor of $4$ along the spectral dimension to restore the original spectral size.

\subsubsection{Random Noise}
\noindent \textbf{Background.} In hyperspectral imaging, captured HSI often contains noise due to environmental interference, lower-quality sensors, or signal transmission issues. Thus, HSI denoising is crucial for HSI restoration and its applications.

\noindent \textbf{Synthesis.} To enhance the diversity of noise magnitude in our dataset, we simulate AWGN using Signal-to-Noise Ratios (SNR) sampled from a normal distribution with mean $\mu$ and standard deviation $\sigma$ to control the noise intensity. The Gaussian noise $I_{\epsilon}$, randomly drawn from $\mathcal{N}(\mathbf{0}, \mathbf{1})$, is scaled by the intensity term to generate the AWGN. The corresponding formula and its effect on the HSI are expressed as:




\begin{equation}
\footnotesize
I_{\text{noisy}} = I_{\text{clean}} + N_{\text{awgn}} =  I_{\text{clean}} + I_{\epsilon} \cdot \sqrt{ \frac{ \| I_{\text{clean}} \|_2^2 }{\text{SNR}}},
\end{equation}

\subsubsection{Band-Missing}
\noindent \textbf{Background.} Due to the conditions of satellite sensors and signal interference during transmission, HSIs often encounter band-missing issues in addition to noise \cite{bandmissing}. 

\noindent \textbf{Synthesis.} We simulate this degradation and categorize it into three types as follows \cite{rtcs}. The first is where the pixel values of certain bands are completely missing, referred to as \ding{182} \textbf{\emph{Band Complete Missing}}. The second and third types involve missing pixel values in certain rows of the image, resulting in stripe-like defects. These are classified as \ding{183} \textbf{\emph{Band-Wise Missing}} and \ding{184} \textbf{\emph{Partial Missing}}, based on whether the rows disappear in a regular or irregular pattern. The visualization of the three different missing types are shown in Figure \ref{fig:bandmissing}.

\begin{table}[htp!]
\centering
\caption{\small \underline{\textbf{Text prompt formats.}} Different colors represent various types of degradation. If a particular degradation type is absent, it will be excluded from the text prompt. The notation [] specifies the subtype of the respective degradation type, while $N_{mb}$ denotes the number of bands affected by band-missing degradation.}
\label{table:text_format}
\scalebox{0.86}{
\begin{tabular}{c|p{6.2cm}} 
\toprule[0.15em]
    \multicolumn{1}{c|}{\textbf{Prompt type}} & \multicolumn{1}{c}{\textbf{Prompt format}} \\ 
    \hline
    \multirow{4}{*}{Long prompt} & "This hyperspectral image \textcolor{Thistle}{faces with '[\textit{band-missing type}]' on $N_{mb}$ bands;} \textcolor{SeaGreen}{it also confronts '[\textit{cloudy type]}'}, \textcolor{BurntOrange}{'[\textit{noisy]}';} \textcolor{Plum}{besides, there exists 'blurring effect in [\textit{blurring type]} domain'}. " \\
    \hline
    \multirow{2}{*}{Short prompt} & "\textcolor{SeaGreen}{[\textit{cloudy type}]}, \textcolor{BurntOrange}{[\textit{noisy}]}, \textcolor{Plum}{[\textit{blurring type}]}, \textcolor{Thistle}{[\textit{band-missing type}]}" \\
\bottomrule[0.15em]
\end{tabular}}
\end{table}

\begin{table}[htp!]
\centering
\caption{\small \underline{\textbf{Effect of text-prompt length.}}}
\label{table:text}
\scalebox{0.8}{
\begin{tabular}{ c| c| c| c}
\toprule[0.15em]
\rowcolor{background_color}
Length & PSNR / SAM / RMSE / ERGAS  & 
\#Params & FLOPs \\\rowcolor{input_color}
\midrule[0.1em]
\rowcolor{input_color}
Long & \secondBest{25.879} / \secondBest{6.684} / \secondBest{0.0196} / \secondBest{10.789} & 26.15M & 135.83G \\
\rowcolor{ours_color}
\textbf{Short} & \best{26.443} / \best{6.060} / \best{0.0186} / \best{8.357} & 26.15M & 135.83G \\
\bottomrule[0.15em]
\end{tabular}}
\end{table}

%% file: sections/architecture_sup.tex
\section {Network Details}
\label{sec:suparch}
\begin{figure*}[htbp]
    \centering
    \includegraphics[width=1.00\textwidth]{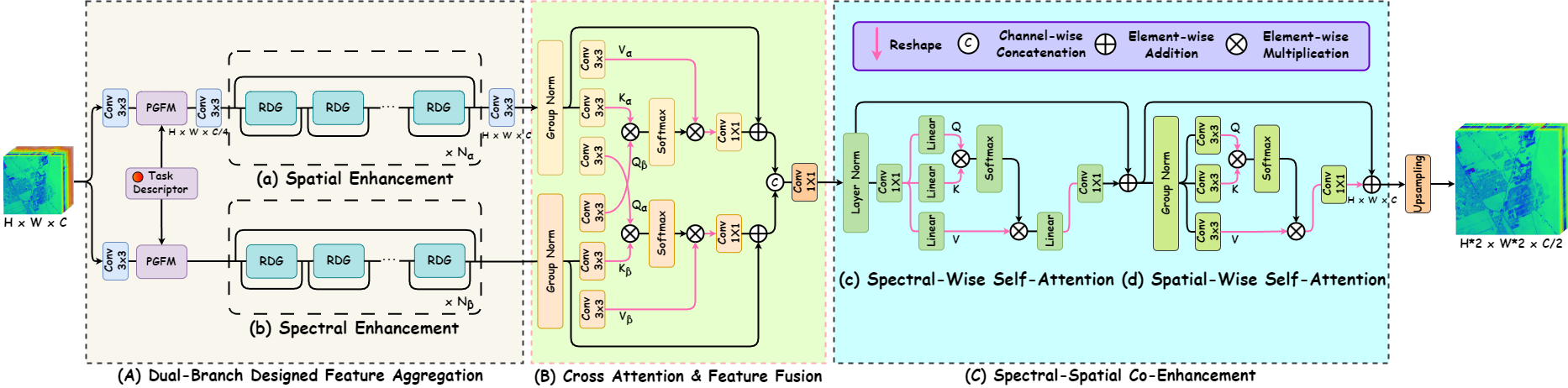}
    \caption{The detailed Architecture of our proposed Prompt-guided Feature Aggregation Block (PGFAB).}
    \label{fig:first}
\end{figure*}

In terms of our PromptHSI architecture, all configurations and output sizes of each layer are displayed in Table \ref{tab:model_arch}. 
Afterwards, we introduce the details of the Prompt-guided Feature Aggregation Block (PGFAB), as showan in Figure \ref{fig:first}, which consists of the Prompt-guided Feature Modulation (PGFM) module and feature enhancement. The motivation of PGFM have been illustrated in our manuscript; here we further discuss the details of feature enhancement.

\noindent \textbf{Dual-branch design for feature aggregation.} Inspired by SPACE \cite{SAMtwobranch}, which uses two branches designed for efficient HSI compressed sensing. This design allows the network to focus independently on learning spectral and spatial features. We adopt the Residual Dense Group (RDG) \cite{DRCT} for the spatial feature extraction branch , and the Fast Residual Dense Block (FRDB) \cite{rtcs} for the spectral feature extraction branch. 


\noindent \textbf{Residual dense group.} This architecture incorporates Swin Transformer \cite{swinir,swintransformer} with dense connections, demonstrating a more lightweight and efficient parameter-size compared to the Residual Hybrid Attention Groups (RHAG) in HAT \cite{HAT}. This design exhibits superior capability in capturing global spatial details.

\noindent \textbf{Fast residual dense block.} Built on group convolutions integrated with dense connections \cite{densenet}, excels at capturing local spatial-spectral details. In particular, the group convolution effectively leverages the low-rank prior \cite{DHP,DPHSIR} inherent in HSI while achieving a reduced parameter-size compared to standard convolutions. It aligns with the HSI characteristic that spectral information exhibits high correlation in local regions while containing substantial redundancy in long-range interactions.



\begin{table*}[htbp]
    \renewcommand{\arraystretch}{1.1}
    \centering
    \caption{\small \underline{\textbf{The configurations of the PromptHSI architecture.}} For some notations used in the Table, $H$, $W$, and $C$ represent the height, width, and the number of bands of the input HSI, respectively. $N_\alpha$ and $N_\beta$ denote the depth of the RDG \cite{DRCT} and FRDB \cite{rtcs} blocks in the dual-branch designed feature aggregation. $c$, $k$, $s$, $head$, $ws$, $ps$, and $n$ express \textbf{the number of output channels}, \textbf{kernel size}, \textbf{stride}, \textbf{attention heads}, \textbf{window size}, \textbf{patch size}, and \textbf{the number of blocks}, respectively. $ds$-$conv$ refers to depthwise separable convolution \cite{dsconv}, $bn$ stands for batch normalization.
    For the task descriptor generation input, the text-prompt for degradation description follows a short format like "$description^1, \ldots, description^n$", as shown in Table \ref{table:text}.}
    \label{tab:model_arch}
        \begin{adjustbox}{max width=\textwidth}
        \begin{tabular}[c]{c|l|l|l} 
        \Xhline{2pt}
        Models & Layers & Configurations & Output size \\
        \hline
        
        & Input1 & Degradation text prompt & 1 string \\
        Task descriptor& Text tokenizer & BPE-based subword text tokenizer & 77 \\
        generation& CLIP text encoder \cite{CLIP} & ViT-B/32 & $512$ \\
        & Adapter & MLP1$(c = 64)$, $leakyrelu$, MLP2$(c = 512)$ & $512$ \\
        \hline
        \hline
        \multirow{18}{*}{PromptHSI} 
        & \multicolumn{3}{c}{Encoder} \\
        \cline{2-4}
        & Input2 & Degraded hyperspectral image & ${H}\times{W}\times{C}$ \\
        & Conv1 & $c = 64, k = 5, s = 1$ & ${H}\times{W}\times{64}$ \\
        & SSAFEB1 & $ds$-$conv(c = 128, k = 5, s = 2), bn, leakyrelu$ & ${\frac{H}{2}}\times{\frac{W}{2}}\times{128}$ \\
        & SSAFEB2 & $ds$-$conv(c = 256, k = 3, s = 2), bn, leakyrelu$ & ${\frac{H}{4}}\times{\frac{W}{4}}\times{256}$ \\
        & SSAFEB3 & $ds$-$conv(c = 512, k = 3, s = 2), bn, leakyrelu$ & ${\frac{H}{8}}\times{\frac{W}{8}}\times{512}$ \\
        \cline{2-4}\cline{2-4}\cline{2-4}
        & \multicolumn{3}{c}{Decoder} \\
        \cline{2-4}
        & PGFAB1 & $c = 256, head = 8, ws = 7, ps = 4, N_\alpha = 2, N_\beta = 1$ & ${\frac{H}{4}}\times{\frac{W}{4}}\times{256}$ \\
        & Concatenation1 & PGFAB1 + SSAFEB2 & ${\frac{H}{4}}\times{\frac{W}{4}}\times{512}$ \\
        & Conv2 & $c = 256, k = 1, s = 1$ & ${\frac{H}{4}}\times{\frac{W}{4}}\times{256}$ \\
        & PGFAB2 & $c = 128, head = 8, ws = 7, ps = 4, N_\alpha = 2, N_\beta = 1$ & ${\frac{H}{2}}\times{\frac{W}{2}}\times{128}$ \\
        & Concatenation2 & PGFAB2 + SSAFEB1 & ${\frac{H}{2}}\times{\frac{W}{2}}\times{256}$ \\
        & Conv3 & $c = 128, k = 1, s = 1$ & ${\frac{H}{2}}\times{\frac{W}{2}}\times{128}$ \\
        & PGFAB3 & $c = D, head = 4, ws = 7, ps = 4, N_\alpha = 2, N_\beta = 1$ & ${H}\times{W}\times{64}$ \\
        & Concatenation3 & PGFAB3 + Conv1 & ${H}\times{W}\times{128}$ \\
        & Conv4 & $c = 128, k = 1, s = 1$ & ${H}\times{W}\times{128}$ \\
        & Transformer blocks & $c = 128, head = 4, n = 2 $ & ${H}\times{W}\times{128}$ \\
        & Conv5 & $c = C, k = 1, s = 1$ & ${H}\times{W}\times{C}$ \\
        \Xhline{2pt}
        \end{tabular}
        \end{adjustbox}
\end{table*}

%% file: sections/experiments_sup.tex
\section {Additional Experiments}
\label{sec:supexperiments}
In addition to the experimental details and ablation studies presented in the manuscript, this section conducts further experiments to evaluate \ding{182} the effects of different encoders, \ding{183} effect of the length of degradation descriptions, and \ding{184} effect of the feature enhancement. Rows are colored to distinguish different approaches: \colorbox{ours_color}{Adopted}, \colorbox{input_color}{Not adopted}.


\subsection{Ablation Study}
We provide additional ablation studies and comparisons with various modules in Table \ref{table:ablation_condition_information}, Table \ref{table:ablation_injection}, and Table \ref{table:ablation_ref} to validate the effectiveness of these modules. 

\noindent \textbf{Effect of text-prompt length.} We further compared the impact of long and short text prompts for the model's performance, where the formats of text-prompts and results are shown in Tables \ref{table:text_format} and \ref{table:text}. The long text-prompt provides distinct descriptions for different degradation types, combining them into a single statement with a tone closer to natural human linguistic speech compared to the short prompt. 


\input{tables/006_appendix_1_2}

\begin{figure*}[htbp]
    \centering
    \includegraphics[width=0.9\textwidth]{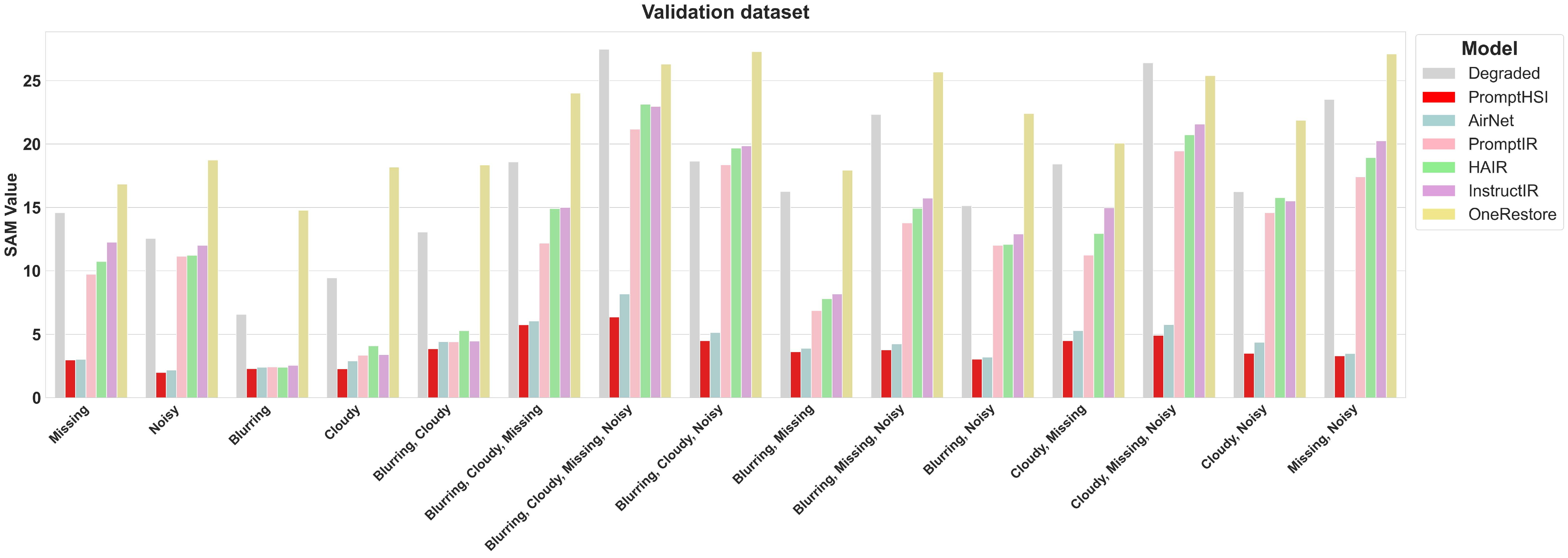}
\vspace*{-4mm}
    \includegraphics[width=0.9\textwidth]{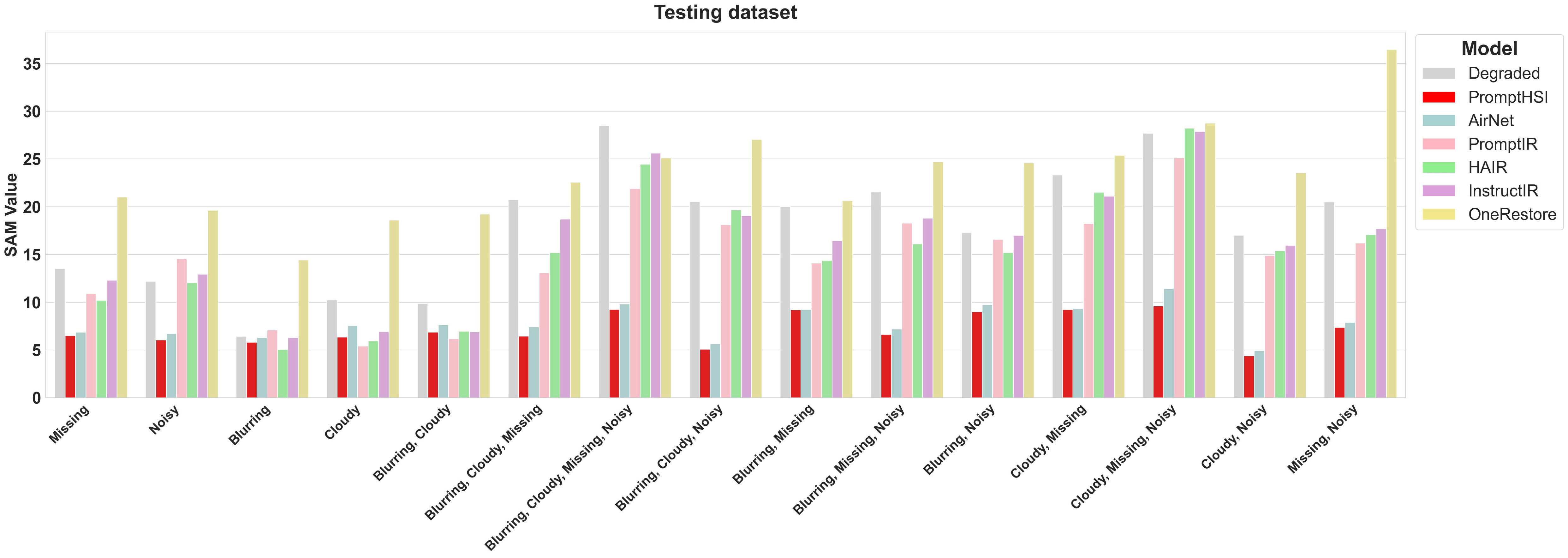}
    \caption{PromptHSI and other All-in-One RGB image restoration methods for different degradations.}
    \label{fig:cross_model_performance}
    
\vspace*{-4mm}
\end{figure*}

\subsection{Visualization Results Analysis}


\noindent \textbf{t-SNE visulization.} As shown in Figure \ref{fig:teaser}, we compare our PromptHSI framework with other All-in-One image restoration methods to identify different degradations. Our approach demonstrates the superior discriminative ability for both multiple and composite degradations within t-SNE visualizations. 
This indicates that our text-prompts and feature modulation mechanism can effectively guide the model in identifying degradation types compared with conventional prompt-based All-in-One RGB image restoration methods.

\begin{figure*}
    \centering
    \includegraphics[width=0.98\linewidth]{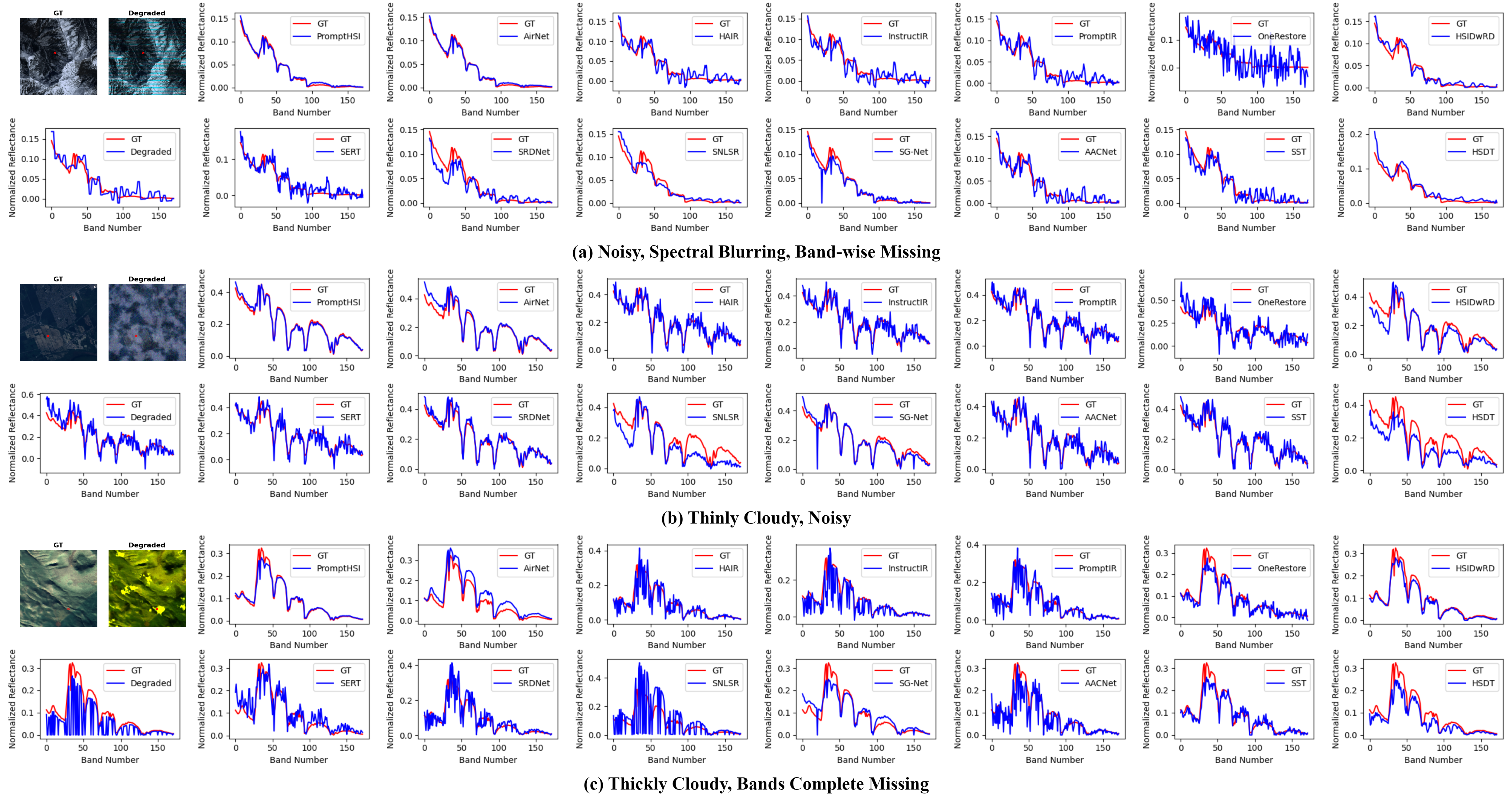}
    \caption{Spectral signatures across 172 bands under three different composite degradation conditions.}
    \label{fig:spectral_signature}
\end{figure*}

\noindent \textbf{Spectral signature analysis.} HSIs are characterized by their spectral signatures, where each substance exhibits a unique spectral profile. Therefore, accurate restoration in spectral domain for degraded HSIs is crucial for downstream applications, such as unmixing or material identification. Figure \ref{fig:spectral_signature} presents the spectral reflectance curves between Ground-truth and reconstructed HSI, demonstrating that the proposed PromptHSI outperforms other methods in these cases.

\begin{figure*}
    \centering
    \includegraphics[width=1.01\linewidth]{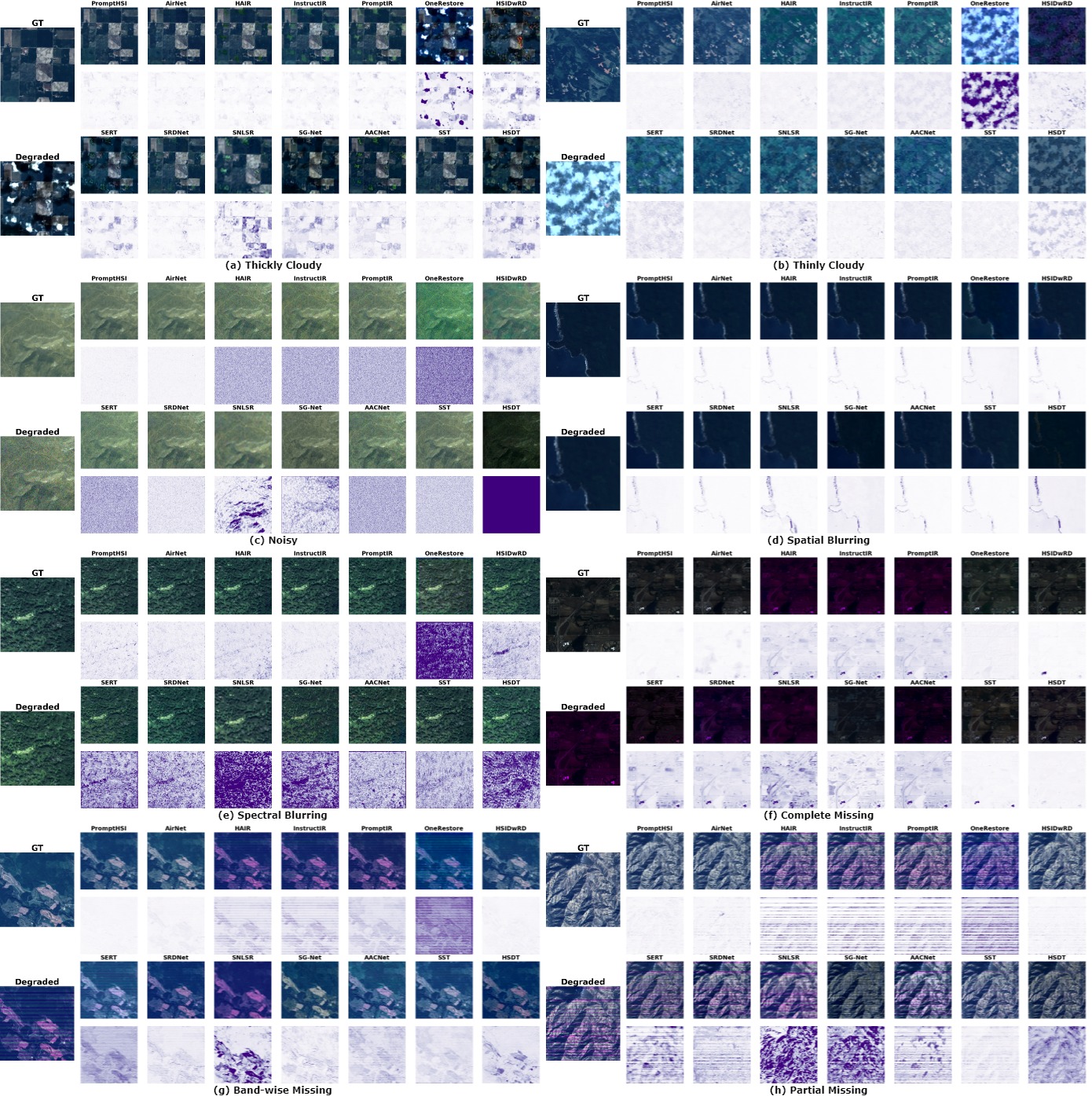}
    \caption{Visual comparison of single-degradation restoration results with peer methods.}
    \label{fig:s1}
\end{figure*}
\begin{figure*}[t]
    \centering
    \includegraphics[width=1.01\linewidth]{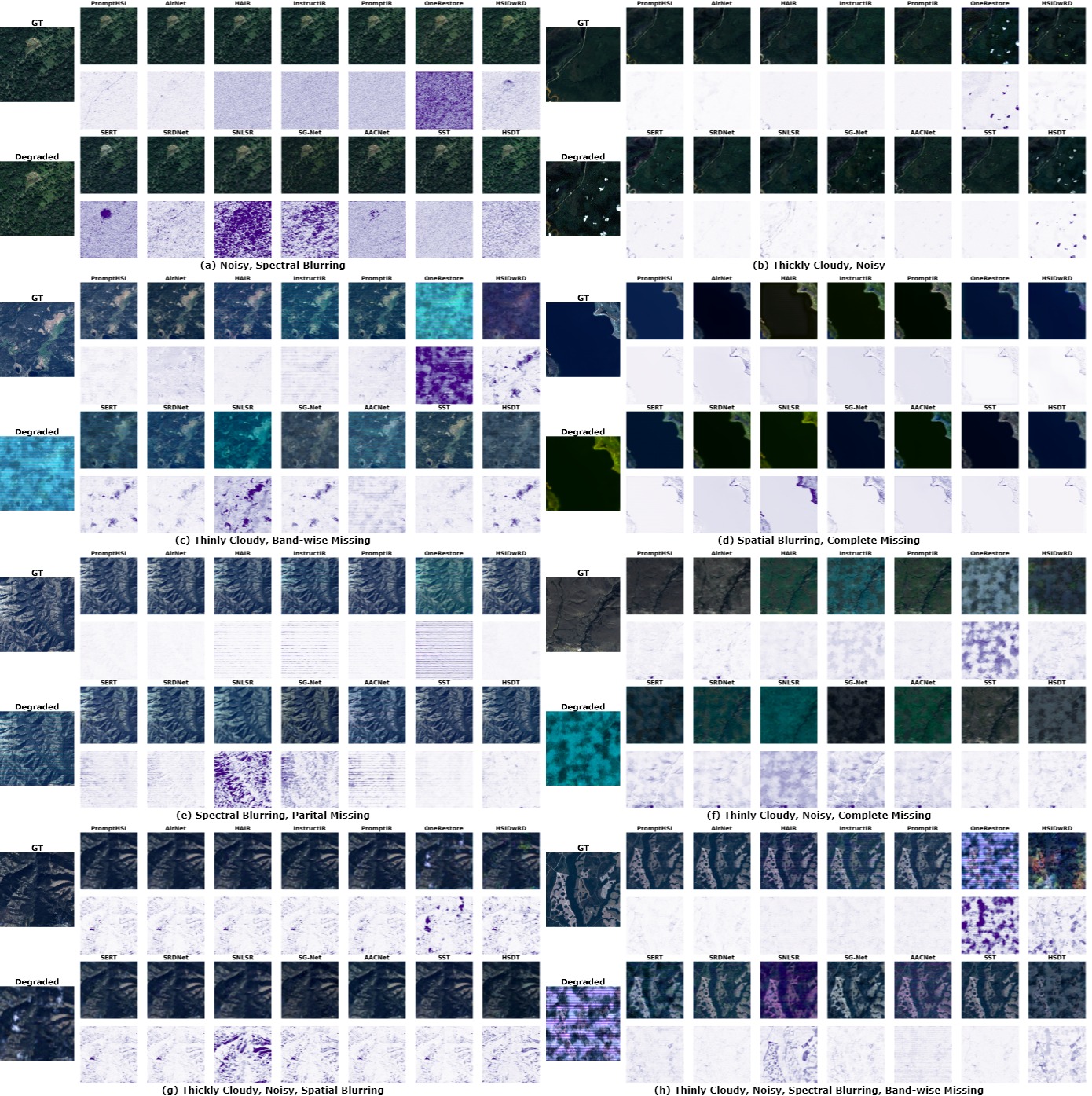}
    \caption{Visual comparison of composite-degradation restoration results with peer methods.
    }
    \label{fig:c1}
\end{figure*}

%% file: tables/006_appendix_1_2.tex
\definecolor{background_color}{RGB}{255,255,255}  

\definecolor{cnn_color}{RGB}{240,248,255}  
\definecolor{transformer_color}{RGB}{255,250,240}  
\definecolor{ours_color}{RGB}{240,255,240}  
\definecolor{input_color}{RGB}{255,240,240}      

\begin{table*}[t]
\parbox{.32\linewidth}{
\centering
\caption{\small Impact of the type of convolution in SSAFEB. Note that S indicate standard convolution; DS represent depthwise separable convolution \cite{dsconv}.}
\label{table:ablation_condition_information}
\vspace{-3mm}
\setlength{\tabcolsep}{2pt}
\scalebox{0.61}{
\begin{tabular}{c|c|c|c}
\toprule[0.15em]
\rowcolor{background_color}
Type &\text{PSNR} / \text{SAM} / \text{RMSE} / \text{ERGAS} & \#Params & FLOPs \\
\rowcolor{input_color}
\hline
S & \secondBest{26.112} / \secondBest{6.307} / \secondBest{0.0186} / \secondBest{10.258} & \secondBest{27.65M} & \secondBest{153.17G} \\ 
\rowcolor{ours_color}
\textbf{DS} & \best{26.443} / \best{6.060} / \best{0.0187} / \best{8.357} & \best{26.15M} & \best{135.83G} \\
\bottomrule[0.15em]
\end{tabular}
}
}
\hfill
\parbox{.34\linewidth}{
\centering
\caption{\small Ablation of feature co-enhancement.}
\label{table:ablation_injection}
\vspace{-3mm}
\setlength{\tabcolsep}{2pt}
\scalebox{0.62}{
\begin{tabular}{cc|c|c|c}
\toprule[0.15em]
%
\rowcolor{background_color}
$\text{Spa}$& $\text{Spe}$ &\text{PSNR} / \text{SAM} / \text{RMSE} / \text{ERGAS} & \#Params & FLOPs \\
\rowcolor{input_color}
\hline
$\times$& $\times$ & 25.846 / 6.403 / 0.0191 / 11.514 & 22.70M & 129.67G \\ \rowcolor{input_color}
$\checkmark$& $\times$ & 26.170 / 6.295 / 0.0187 / 12.911 & 24.08M  & 132.14G  \\ \rowcolor{input_color}
$\times$& $\checkmark$ & {25.247} / {6.678} / {0.0196} / {10.582} & 24.77M & 133.37G \\\hline
\rowcolor{ours_color}
\textbf{$\checkmark$}&\textbf{$\checkmark$} & \textbf{26.443} / \textbf{6.060} / \textbf{0.0186} / \textbf{8.357} & 26.15M & 135.83G \\
\bottomrule[0.15em]
\end{tabular}
}}
\hfill
\parbox{.32\linewidth}{
\centering
\caption{\small Ablation of attention in PGFAB.}
\label{table:ablation_ref}
\vspace{-3mm}
\setlength{\tabcolsep}{4pt}
\scalebox{0.62}{
\begin{tabular}{c|c|c|c}
\toprule[0.15em]
\rowcolor{background_color}
%
\rowcolor{background_color}
$\text{Without}$ &\text{PSNR} / \text{SAM} / \text{RMSE} / \text{ERGAS} & \#Params & FLOPs \\
\rowcolor{ours_color}
\hline
\textbf{Full model} & \textbf{26.443} / \textbf{6.060} / \textbf{0.0186} / \textbf{8.357} & 26.15M & 135.83G \\ \hline\rowcolor{input_color}
{Spat (RDG)} & 25.330 / 6.595 / 0.0196 / 9.894 & 14.37M & 106.49G \\ 
\rowcolor{input_color}
{Spec (FRDB)} & 26.475 / 6.224 / 0.0182 / 9.812 & 18.96M & 122.12G \\
\rowcolor{input_color}
{Spat $\&$ Spec} & 25.403 / 6.575 / 0.0196 / 10.628 & 5.64M & 96.48G \\
\bottomrule[0.15em]
\end{tabular}
}}
\vspace*{-1mm}
\end{table*}

%% file: main.bbl
\begin{thebibliography}{63}
\providecommand{\natexlab}[1]{#1}
\providecommand{\url}[1]{\texttt{#1}}
\expandafter\ifx\csname urlstyle\endcsname\relax
  \providecommand{\doi}[1]{doi: #1}\else
  \providecommand{\doi}{doi: \begingroup \urlstyle{rm}\Url}\fi

\bibitem[Bioucas-Dias et~al.(2013)Bioucas-Dias, Plaza, Camps-Valls, Scheunders, Nasrabadi, and Chanussot]{HSI}
Jose~M. Bioucas-Dias, Antonio Plaza, Gustavo Camps-Valls, Paul Scheunders, Nasser Nasrabadi, and Jocelyn Chanussot.
\newblock Hyperspectral remote sensing data analysis and future challenges.
\newblock \emph{IEEE Geoscience and Remote Sensing Magazine}, 1\penalty0 (2):\penalty0 6--36, 2013.

\bibitem[Cai et~al.(2022{\natexlab{a}})Cai, Lin, Hu, Wang, Yuan, Zhang, Timofte, and Gool]{mst}
Yuanhao Cai, Jing Lin, Xiaowan Hu, Haoqian Wang, Xin Yuan, Yulun Zhang, Radu Timofte, and Luc~Van Gool.
\newblock Mask-guided spectral-wise transformer for efficient hyperspectral image reconstruction.
\newblock In \emph{CVPR}, 2022{\natexlab{a}}.

\bibitem[Cai et~al.(2022{\natexlab{b}})Cai, Lin, Lin, Wang, Zhang, Pfister, Timofte, and Gool]{mst_pp}
Yuanhao Cai, Jing Lin, Zudi Lin, Haoqian Wang, Yulun Zhang, Hanspeter Pfister, Radu Timofte, and Luc~Van Gool.
\newblock Mst++: Multi-stage spectral-wise transformer for efficient spectral reconstruction.
\newblock In \emph{CVPRW}, 2022{\natexlab{b}}.

\bibitem[Cao et~al.(2024)Cao, Cao, Pang, Meng, and Cao]{HAIR}
Jin Cao, Yi Cao, Li Pang, Deyu Meng, and Xiangyong Cao.
\newblock Hair: Hypernetworks-based all-in-one image restoration, 2024.

\bibitem[Cerra et~al.(2014)Cerra, Müller, and Reinartz]{UBD}
Daniele Cerra, Rupert Müller, and Peter Reinartz.
\newblock Unmixing-based denoising for destriping and inpainting of hyperspectral images.
\newblock In \emph{2014 IEEE Geoscience and Remote Sensing Symposium}, pages 4620--4623, 2014.

\bibitem[Chen et~al.(2022)Chen, Huang, Tsai, Yang, Ding, and Kuo]{TKMANet}
Wei-Ting Chen, Zhi-Kai Huang, Cheng-Che Tsai, Hao-Hsiang Yang, Jian-Jiun Ding, and Sy-Yen Kuo.
\newblock Learning multiple adverse weather removal via two-stage knowledge learning and multi-contrastive regularization: Toward a unified model.
\newblock In \emph{2022 IEEE/CVF Conference on Computer Vision and Pattern Recognition (CVPR)}, pages 17632--17641, 2022.

\bibitem[Chen et~al.(2023)Chen, Wang, Zhou, Qiao, and Dong]{HAT}
Xiangyu Chen, Xintao Wang, Jiantao Zhou, Yu Qiao, and Chao Dong.
\newblock Activating more pixels in image super-resolution transformer.
\newblock In \emph{Proceedings of the IEEE/CVF Conference on Computer Vision and Pattern Recognition (CVPR)}, pages 22367--22377, 2023.

\bibitem[Chen et~al.(2024)Chen, Li, Pu, Liu, Zhou, Qiao, and Dong]{Xrestormer}
Xiangyu Chen, Zheyuan Li, Yuandong Pu, Yihao Liu, Jiantao Zhou, Yu Qiao, and Chao Dong.
\newblock A comparative study of image restoration networks for general backbone network design.
\newblock In \emph{European Conference on Computer Vision (ECCV)}, 2024.

\bibitem[Chollet(2016)]{dsconv}
Fran{\c{c}}ois Chollet.
\newblock Xception: Deep learning with depthwise separable convolutions.
\newblock \emph{CoRR}, abs/1610.02357, 2016.

\bibitem[Conde et~al.(2024)Conde, Geigle, and Timofte]{InstructIR}
Marcos~V Conde, Gregor Geigle, and Radu Timofte.
\newblock Instructir: High-quality image restoration following human instructions.
\newblock In \emph{Proceedings of the European Conference on Computer Vision (ECCV)}, 2024.

\bibitem[Cui et~al.(2024)Cui, Zamir, Khan, Knoll, Shah, and Khan]{adair}
Yuning Cui, Syed~Waqas Zamir, Salman Khan, Alois Knoll, Mubarak Shah, and Fahad~Shahbaz Khan.
\newblock Adair: Adaptive all-in-one image restoration via frequency mining and modulation, 2024.

\bibitem[Czerkawski et~al.(2023)Czerkawski, Atkinson, Michie, and Tachtatzis]{cloudgenerator}
Mikolaj Czerkawski, Robert Atkinson, Craig Michie, and Christos Tachtatzis.
\newblock Satellitecloudgenerator: Controllable cloud and shadow synthesis for multi-spectral optical satellite images.
\newblock \emph{Remote Sensing}, 15\penalty0 (17), 2023.

\bibitem[Faghih~Niresi and Chi(2023)]{Prior1}
Keivan Faghih~Niresi and Chong-Yung Chi.
\newblock Robust hyperspectral inpainting via low-rank regularized untrained convolutional neural network.
\newblock \emph{IEEE Geoscience and Remote Sensing Letters}, 20:\penalty0 1--5, 2023.

\bibitem[Fan et~al.(2021)Fan, Chen, Yuan, Hua, Yu, and Chen]{DL}
Qingnan Fan, Dongdong Chen, Lu Yuan, Gang Hua, Nenghai Yu, and Baoquan Chen.
\newblock A general decoupled learning framework for parameterized image operators.
\newblock \emph{IEEE Transactions on Pattern Analysis and Machine Intelligence}, 43\penalty0 (1):\penalty0 33--47, 2021.

\bibitem[Guo et~al.(2024)Guo, Gao, Lu, Liu, and He]{OneRestore}
Yu Guo, Yuan Gao, Yuxu Lu, Ryan~Wen Liu, and Shengfeng He.
\newblock Onerestore: A universal restoration framework for composite degradation.
\newblock In \emph{European Conference on Computer Vision (ECCV)}, 2024.

\bibitem[He et~al.(2023)He, Yuan, Li, Xiao, Liu, Shen, and Zhang]{SSR-dl}
Jiang He, Qiangqiang Yuan, Jie Li, Yi Xiao, Denghong Liu, Huanfeng Shen, and Liangpei Zhang.
\newblock Spectral super-resolution meets deep learning: Achievements and challenges.
\newblock \emph{Information Fusion}, 97:\penalty0 101812, 2023.

\bibitem[He et~al.(2019)He, Yao, Li, Yokoya, and Zhao]{NGmeet}
Wei He, Quanming Yao, Chao Li, Naoto Yokoya, and Qibin Zhao.
\newblock Non-local meets global: An integrated paradigm for hyperspectral denoising.
\newblock In \emph{2019 IEEE/CVF Conference on Computer Vision and Pattern Recognition (CVPR)}, pages 6861--6870, 2019.

\bibitem[Hsu et~al.(2024{\natexlab{a}})Hsu, Jian, Tu, Lee, and Chen]{rtcs}
Chih-Chung Hsu, Chih-Yu Jian, Eng-Shen Tu, Chia-Ming Lee, and Guan-Lin Chen.
\newblock Real-time compressed sensing for joint hyperspectral image transmission and restoration for cubesat.
\newblock \emph{IEEE Transactions on Geoscience and Remote Sensing}, 2024{\natexlab{a}}.

\bibitem[Hsu et~al.(2024{\natexlab{b}})Hsu, Lee, and Chou]{DRCT}
Chih-Chung Hsu, Chia-Ming Lee, and Yi-Shiuan Chou.
\newblock Drct: Saving image super-resolution away from information bottleneck.
\newblock In \emph{Proceedings of the IEEE/CVF Conference on Computer Vision and Pattern Recognition (CVPR) Workshops}, pages 6133--6142, 2024{\natexlab{b}}.

\bibitem[Hsu et~al.(2024{\natexlab{c}})Hsu, Ni, Lee, and Kang]{CSAKD}
Chih-Chung Hsu, Chih-Chien Ni, Chia-Ming Lee, and Li-Wei Kang.
\newblock Csakd: Knowledge distillation with cross self-attention for hyperspectral and multispectral image fusion.
\newblock \emph{arXiv preprint arXiv:2406.19666}, 2024{\natexlab{c}}.

\bibitem[Hu et~al.(2024)Hu, Wang, Jiang, Zhang, and Ma]{SNLSR}
Qian Hu, Xinya Wang, Junjun Jiang, Xiao-Ping Zhang, and Jiayi Ma.
\newblock Exploring the spectral prior for hyperspectral image super-resolution.
\newblock \emph{IEEE Transactions on Image Processing}, 33:\penalty0 5260--5272, 2024.

\bibitem[Huang et~al.(2017)Huang, Liu, Van Der~Maaten, and Weinberger]{densenet}
Gao Huang, Zhuang Liu, Laurens Van Der~Maaten, and Kilian~Q. Weinberger.
\newblock Densely connected convolutional networks.
\newblock In \emph{2017 IEEE Conference on Computer Vision and Pattern Recognition (CVPR)}, pages 2261--2269, 2017.

\bibitem[Jiang et~al.(2020)Jiang, Sun, Liu, and Ma]{SSPSR}
Junjun Jiang, He Sun, Xianming Liu, and Jiayi Ma.
\newblock Learning spatial-spectral prior for super-resolution of hyperspectral imagery.
\newblock \emph{IEEE Transactions on Computational Imaging}, 6:\penalty0 1082--1096, 2020.

\bibitem[Lai et~al.(2023{\natexlab{a}})Lai, Lin, and Leng]{HyperRestormer}
Yo-Yu Lai, Chia-Hsiang Lin, and Zi-Chao Leng.
\newblock Hyper-restormer: A general hyperspectral image restoration transformer for remote sensing imaging, 2023{\natexlab{a}}.

\bibitem[Lai et~al.(2022)Lai, Wei, and Fu]{DPHSIR}
Zeqiang Lai, Kaixuan Wei, and Ying Fu.
\newblock Deep plug-and-play prior for hyperspectral image restoration.
\newblock \emph{Neurocomputing}, 481:\penalty0 281--293, 2022.

\bibitem[Lai et~al.(2023{\natexlab{b}})Lai, Chenggang, and Fu]{HSDT}
Zeqiang Lai, Yan Chenggang, and Ying Fu.
\newblock Hybrid spectral denoising transformer with guided attention.
\newblock In \emph{Proceedings of the IEEE International Conference on Computer Vision}, 2023{\natexlab{b}}.

\bibitem[Li et~al.(2022)Li, Liu, Hu, Wu, Lv, and Peng]{AirNet}
Boyun Li, Xiao Liu, Peng Hu, Zhongqin Wu, Jiancheng Lv, and Xi Peng.
\newblock {All-In-One Image Restoration for Unknown Corruption}.
\newblock In \emph{IEEE Conference on Computer Vision and Pattern Recognition}, New Orleans, LA, 2022.

\bibitem[Li et~al.(2023{\natexlab{a}})Li, Fu, and Zhang]{SST}
Miaoyu Li, Ying Fu, and Yulun Zhang.
\newblock Spatial-spectral transformer for hyperspectral image denoising.
\newblock In \emph{AAAI}, 2023{\natexlab{a}}.

\bibitem[Li et~al.(2023{\natexlab{b}})Li, Liu, Fu, Zhang, and Dou]{SERT}
Miaoyu Li, Ji Liu, Ying Fu, Yulun Zhang, and Dejing Dou.
\newblock Spectral enhanced rectangle transformer for hyperspectral image denoising.
\newblock In \emph{CVPR}, 2023{\natexlab{b}}.

\bibitem[Li et~al.(2018)Li, Zhang, Dingl, Wei, and Zhang]{GDRRN}
Yong Li, Lei Zhang, Chen Dingl, Wei Wei, and Yanning Zhang.
\newblock Single hyperspectral image super-resolution with grouped deep recursive residual network.
\newblock In \emph{2018 IEEE Fourth International Conference on Multimedia Big Data (BigMM)}, pages 1--4, 2018.

\bibitem[Liang et~al.(2021)Liang, Cao, Sun, Zhang, Van~Gool, and Timofte]{swinir}
Jingyun Liang, Jiezhang Cao, Guolei Sun, Kai Zhang, Luc Van~Gool, and Radu Timofte.
\newblock Swinir: Image restoration using swin transformer.
\newblock \emph{arXiv preprint arXiv:2108.10257}, 2021.

\bibitem[Lin et~al.(2020)Lin, Bioucas~Dias, Lin, Lin, and Kao]{SAMtwobranch}
Chia-Hsiang Lin, Jose~M. Bioucas~Dias, Tzu-Hsuan Lin, Yen-Cheng Lin, and Chi-Hung Kao.
\newblock A new hyperspectral compressed sensing method for efficient satellite communications.
\newblock In \emph{2020 IEEE 11th Sensor Array and Multichannel Signal Processing Workshop (SAM)}, pages 1--5, 2020.

\bibitem[Lin et~al.(2022{\natexlab{a}})Lin, Lin, Lin, and Lin]{FR}
Chia-Hsiang Lin, Tzu-Hsuan Lin, Ting-Hsuan Lin, and Tang-Huang Lin.
\newblock Fast reconstruction of hyperspectral image from its rgb counterpart using admm-adam theory.
\newblock In \emph{2022 12th Workshop on Hyperspectral Imaging and Signal Processing: Evolution in Remote Sensing (WHISPERS)}, pages 1--5, 2022{\natexlab{a}}.

\bibitem[Lin et~al.(2022{\natexlab{b}})Lin, Lin, and Tang]{ADMMADAM}
Chia-Hsiang Lin, Yen-Cheng Lin, and Po-Wei Tang.
\newblock Admm-adam: A new inverse imaging framework blending the advantages of convex optimization and deep learning.
\newblock \emph{IEEE Transactions on Geoscience and Remote Sensing}, 60:\penalty0 1--16, 2022{\natexlab{b}}.

\bibitem[Lin et~al.(2024{\natexlab{a}})Lin, Hsu, Young, Hsieh, and Tai]{QRCODE}
Chia-Hsiang Lin, Chih-Chung Hsu, Si-Sheng Young, Cheng-Ying Hsieh, and Shen-Chieh Tai.
\newblock Qrcode: Quasi-residual convex deep network for fusing misaligned hyperspectral and multispectral images.
\newblock \emph{IEEE Transactions on Geoscience and Remote Sensing}, 62:\penalty0 1--15, 2024{\natexlab{a}}.

\bibitem[Lin et~al.(2024{\natexlab{b}})Lin, Zhang, Wei, Ren, Jiang, Tian, and Zuo]{TextualDegRemoval}
Jingbo Lin, Zhilu Zhang, Yuxiang Wei, Dongwei Ren, Dongsheng Jiang, Qi Tian, and Wangmeng Zuo.
\newblock Improving image restoration through removing degradations in textual representations.
\newblock In \emph{2024 IEEE/CVF Conference on Computer Vision and Pattern Recognition (CVPR)}, pages 2866--2878, 2024{\natexlab{b}}.

\bibitem[Lin et~al.(2024{\natexlab{c}})Lin, He, Chen, Lyu, Dai, Yu, Ouyang, Qiao, and Dong]{DiffBIR}
Xinqi Lin, Jingwen He, Ziyan Chen, Zhaoyang Lyu, Bo Dai, Fanghua Yu, Wanli Ouyang, Yu Qiao, and Chao Dong.
\newblock Diffbir: Towards blind image restoration with generative diffusion prior, 2024{\natexlab{c}}.

\bibitem[Lin et~al.(2024{\natexlab{d}})Lin, Cheng, Qiu, Kang, Lee, and Hsu]{lin2024selfsupervisedfusariumheadblight}
Yu-Fan Lin, Ching-Heng Cheng, Bo-Cheng Qiu, Cheng-Jun Kang, Chia-Ming Lee, and Chih-Chung Hsu.
\newblock Self-supervised fusarium head blight detection with hyperspectral image and feature mining.
\newblock \emph{arXiv preprint arXiv:2409.00395}, 2024{\natexlab{d}}.

\bibitem[Liu et~al.(2024)Liu, Liu, Zhang, Yuan, Sui, and Chen]{SRDNet}
Tingting Liu, Yuan Liu, Chuncheng Zhang, Liyin Yuan, Xiubao Sui, and Qian Chen.
\newblock Hyperspectral image super-resolution via dual-domain network based on hybrid convolution.
\newblock \emph{IEEE Transactions on Geoscience and Remote Sensing}, 62:\penalty0 1--18, 2024.

\bibitem[Liu et~al.(2021)Liu, Lin, Cao, Hu, Wei, Zhang, Lin, and Guo]{swintransformer}
Ze Liu, Yutong Lin, Yue Cao, Han Hu, Yixuan Wei, Zheng Zhang, Stephen Lin, and Baining Guo.
\newblock Swin transformer: Hierarchical vision transformer using shifted windows.
\newblock In \emph{Proceedings of the IEEE/CVF International Conference on Computer Vision (ICCV)}, 2021.

\bibitem[Loshchilov and Hutter(2019)]{AdamW}
Ilya Loshchilov and Frank Hutter.
\newblock Decoupled weight decay regularization, 2019.

\bibitem[Luo et~al.(2024)Luo, Zhao, Li, Ng, and Meng]{10354352}
Yisi Luo, Xile Zhao, Zhemin Li, Michael~K. Ng, and Deyu Meng.
\newblock Low-rank tensor function representation for multi-dimensional data recovery.
\newblock \emph{IEEE Transactions on Pattern Analysis and Machine Intelligence}, 46\penalty0 (5):\penalty0 3351--3369, 2024.

\bibitem[Ma et~al.(2022)Ma, Wang, and Tong]{SGNET}
Xiaofeng Ma, Qunming Wang, and Xiaohua Tong.
\newblock A spectral grouping-based deep learning model for haze removal of hyperspectral images.
\newblock \emph{ISPRS Journal of Photogrammetry and Remote Sensing}, 188:\penalty0 177--189, 2022.

\bibitem[Maggioni et~al.(2013)Maggioni, Katkovnik, Egiazarian, and Foi]{BM4D}
Matteo Maggioni, Vladimir Katkovnik, Karen Egiazarian, and Alessandro Foi.
\newblock Nonlocal transform-domain filter for volumetric data denoising and reconstruction.
\newblock \emph{IEEE Transactions on Image Processing}, 22\penalty0 (1):\penalty0 119--133, 2013.

\bibitem[Nakhostin et~al.(2016)Nakhostin, Clenet, Corpetti, and Courty]{HSI2}
Sina Nakhostin, Harold Clenet, Thomas Corpetti, and Nicolas Courty.
\newblock Joint anomaly detection and spectral unmixing for planetary hyperspectral images.
\newblock \emph{IEEE Transactions on Geoscience and Remote Sensing}, 54\penalty0 (12):\penalty0 6879--6894, 2016.

\bibitem[Peng et~al.(2022)Peng, Wang, Cao, Liu, Rui, and Meng]{RCTV}
Jiangjun Peng, Hailin Wang, Xiangyong Cao, Xinling Liu, Xiangyu Rui, and Deyu Meng.
\newblock Fast noise removal in hyperspectral images via representative coefficient total variation.
\newblock \emph{IEEE Transactions on Geoscience and Remote Sensing}, 60:\penalty0 1--17, 2022.

\bibitem[Potlapalli et~al.(2023)Potlapalli, Zamir, Khan, and Khan]{PromptIR}
Vaishnav Potlapalli, Syed~Waqas Zamir, Salman Khan, and Fahad Khan.
\newblock Promptir: Prompting for all-in-one image restoration.
\newblock In \emph{Thirty-seventh Conference on Neural Information Processing Systems}, 2023.

\bibitem[Radford et~al.(2021)Radford, Kim, Hallacy, Ramesh, Goh, Agarwal, Sastry, Askell, Mishkin, Clark, et~al.]{CLIP}
Alec Radford, Jong~Wook Kim, Chris Hallacy, Aditya Ramesh, Gabriel Goh, Sandhini Agarwal, Girish Sastry, Amanda Askell, Pamela Mishkin, Jack Clark, et~al.
\newblock Learning transferable visual models from natural language supervision.
\newblock \emph{arXiv preprint arXiv:2103.00020}, 2021.

\bibitem[Shi et~al.(2015)Shi, Cheng, Wang, Yap, and Shen]{LRTV}
Feng Shi, Jian Cheng, Li Wang, Pew-Thian Yap, and Dinggang Shen.
\newblock Lrtv: Mr image super-resolution with low-rank and total variation regularizations.
\newblock \emph{IEEE Transactions on Medical Imaging}, 34\penalty0 (12):\penalty0 2459--2466, 2015.

\bibitem[Sidorov and Yngve~Hardeberg(2019)]{DHP}
Oleksii Sidorov and Jon Yngve~Hardeberg.
\newblock Deep hyperspectral prior: Single-image denoising, inpainting, super-resolution.
\newblock In \emph{Proceedings of the IEEE/CVF International Conference on Computer Vision (ICCV) Workshops}, 2019.

\bibitem[Skakun(2022)]{clouddetector}
Sergii et~al. Skakun.
\newblock Cloud mask intercomparison exercise (cmix): An evaluation of cloud masking algorithms for landsat 8 and sentinel-2.
\newblock \emph{Remote Sensing of Environment}, 274:\penalty0 112990, 2022.

\bibitem[Vane et~al.(1993)Vane, Green, Chrien, Enmark, Hansen, and Porter]{35}
Gregg Vane, Robert Green, Thomas Chrien, Harry Enmark, Earl Hansen, and Wallace Porter.
\newblock The airborne visible/infrared imaging spectrometer (aviris).
\newblock \emph{Remote Sensing of Environment}, 44\penalty0 (2-3):\penalty0 127--143, 1993.

\bibitem[Wang et~al.(2022)Wang, Cun, Bao, Zhou, Liu, and Li]{Uformer}
Zhendong Wang, Xiaodong Cun, Jianmin Bao, Wengang Zhou, Jianzhuang Liu, and Houqiang Li.
\newblock Uformer: A general u-shaped transformer for image restoration.
\newblock In \emph{Proceedings of the IEEE/CVF Conference on Computer Vision and Pattern Recognition (CVPR)}, pages 17683--17693, 2022.

\bibitem[Waqas~Zamir et~al.(2022)Waqas~Zamir, Arora, Khan, Hayat, Shahbaz~Khan, and Yang]{restormer}
Syed Waqas~Zamir, Aditya Arora, Salman Khan, Munawar Hayat, Fahad Shahbaz~Khan, and Ming-Hsuan Yang.
\newblock Restormer: Efficient transformer for high-resolution image restoration.
\newblock In \emph{CVPR}, 2022.

\bibitem[Wu et~al.(2024)Wu, Li, Xu, Huang, and Hoi]{RUN}
Zhijian Wu, Jun Li, Chang Xu, Dingjiang Huang, and Steven C.~H. Hoi.
\newblock Run: Rethinking the unet architecture for efficient image restoration.
\newblock \emph{IEEE Transactions on Multimedia}, 26:\penalty0 10381--10394, 2024.

\bibitem[Xu et~al.(2023)Xu, Peng, Zhang, Jia, and Jia]{AACNet}
Meng Xu, Yanxin Peng, Ying Zhang, Xiuping Jia, and Sen Jia.
\newblock Aacnet: Asymmetric attention convolution network for hyperspectral image dehazing.
\newblock \emph{IEEE Transactions on Geoscience and Remote Sensing}, 61:\penalty0 1--14, 2023.

\bibitem[Yi et~al.(2024)Yi, Xu, Zhang, Tang, and Ma]{textif}
Xunpeng Yi, Han Xu, Hao Zhang, Linfeng Tang, and Jiayi Ma.
\newblock Text-if: Leveraging semantic text guidance for degradation-aware and interactive image fusion.
\newblock In \emph{Proceedings of the IEEE/CVF Conference on Computer Vision and Pattern Recognition (CVPR)}, 2024.

\bibitem[Yuan et~al.(2019)Yuan, Zhang, Li, Shen, and Zhang]{8454887}
Qiangqiang Yuan, Qiang Zhang, Jie Li, Huanfeng Shen, and Liangpei Zhang.
\newblock Hyperspectral image denoising employing a spatial–spectral deep residual convolutional neural network.
\newblock \emph{IEEE Transactions on Geoscience and Remote Sensing}, 57\penalty0 (2):\penalty0 1205--1218, 2019.

\bibitem[Zhang et~al.(2018{\natexlab{a}})Zhang, Yuan, Zeng, Li, and Wei]{bandmissing}
Qiang Zhang, Qiangqiang Yuan, Chao Zeng, Xinghua Li, and Yancong Wei.
\newblock Missing data reconstruction in remote sensing image with a unified spatial–temporal–spectral deep convolutional neural network.
\newblock \emph{IEEE Transactions on Geoscience and Remote Sensing}, 56\penalty0 (8):\penalty0 4274--4288, 2018{\natexlab{a}}.

\bibitem[Zhang et~al.(2021)Zhang, Fu, and Li]{HSIDwRD}
Tao Zhang, Ying Fu, and Cheng Li.
\newblock Hyperspectral image denoising with realistic data.
\newblock In \emph{Proceedings of the IEEE/CVF International Conference on Computer Vision (ICCV)}, pages 2248--2257, 2021.

\bibitem[Zhang et~al.(2022)Zhang, Shi, Liu, Dong, and Wu]{GDM}
Wenlong Zhang, Guangyuan Shi, Yihao Liu, Chao Dong, and Xiao-Ming Wu.
\newblock A closer look at blind super-resolution: Degradation models, baselines, and performance upper bounds.
\newblock In \emph{Proceedings of the IEEE/CVF Conference on Computer Vision and Pattern Recognition (CVPR) Workshops}, pages 527--536, 2022.

\bibitem[Zhang et~al.(2018{\natexlab{b}})Zhang, Li, Li, Wang, Zhong, and Fu]{RCAN}
Yulun Zhang, Kunpeng Li, Kai Li, Lichen Wang, Bineng Zhong, and Yun Fu.
\newblock Image super-resolution using very deep residual channel attention networks.
\newblock In \emph{ECCV}, 2018{\natexlab{b}}.

\bibitem[Zhuang and Bioucas-Dias(2018)]{Fasthyin}
Lina Zhuang and José~M. Bioucas-Dias.
\newblock Fast hyperspectral image denoising and inpainting based on low-rank and sparse representations.
\newblock \emph{IEEE Journal of Selected Topics in Applied Earth Observations and Remote Sensing}, 11\penalty0 (3):\penalty0 730--742, 2018.

\end{thebibliography}
